\documentclass[a4paper,11pt]{article}
\pdfoutput=1
\usepackage{jheppub}


\usepackage{latexsym,amsbsy}
\usepackage[compat=1.1.0]{tikz-feynman}

\newcommand{\BE}{\begin{equation}}
\newcommand{\EE}{\end{equation}}
\newcommand{\BQ}{\begin{equation} \begin{array}{c}}
\newcommand{\EQ}{\end{array}\end{equation}}
\newcommand{\BQA}{\begin{equation} \begin{aligned}}
\newcommand{\EQA}{\end{aligned}\end{equation}}

\newcommand{\gcpl}{\text{g}}
\newcommand{\PhiB}{\overline{\Phi}}
\newcommand{\LX}{\Lambda}

\newcommand{\lX}{\lambda}
\newcommand{\lXB}{\overline{\lX}}
\newcommand{\LAG}{\mathcal{L}}
\newcommand{\sd}{{\sqrt 2}}
\newcommand{\st}{{\sqrt 3}}
\newcommand{\SG}{{\sigma}}
\newcommand{\SB}{{\overline{\sigma}}}

\newcommand{\nn}{\nonumber}
\newcommand{\enl}{\enlargethispage*{\baselineskip}}

\newcommand{\U}{{\rm U}}
\newcommand{\SU}{{\rm SU}}

\newcommand{\SO}{{\rm SO}}
\newcommand{\Sp}{{\rm Sp}}
\newcommand{\OSp}{{\rm OSp}}

\providecommand\inspire[1]{\href{https://inspirehep.net/search?p=find+#1}{{\tiny IN}{\footnotesize SPIRE}}}

\providecommand\erratum[4][ibid.\ ]{\emph{Erratum #1}{\bf #2} (#3) #4}

\providecommand{\jhep}[3] {\ifnum#2>2009%
\href{https://doi.org/10.1007/JHEP#1(#2)#3}{\emph{JHEP} {\bf #1} (#2) #3}%
\else%
\href{https://doi.org/10.1088/1126-6708/#2/#1/#3}{\emph{JHEP} {\bf #1} (#2) #3}%
\fi}
\def\issueFromCounter.#1#2#3#4#5#6.{#2#3}
\providecommand{\jstat}[2]{\PackageWarningNoLine{\jname}{The macro \protect\jstat\space is obsolete!\MessageBreak Please typeset JSTAT as any other journal}%
  \href{https://doi.org/10.1088/1742-5468/#1/\issueFromCounter.#2./#2}{\emph{J.\ Stat.\ Mech.\ }(#1) #2}} 
\providecommand{\hepth}[1]{\href{https://arxiv.org/abs/hep-th/#1}{\tt hep-th/#1}}
\providecommand{\hepph}[1]{\href{https://arxiv.org/abs/hep-ph/#1}{\tt hep-ph/#1}}

\providecommand{\arXivid}[1]{\href{https://arxiv.org/abs/#1}{\tt arXiv:#1}}
\providecommand{\Math}[2]{%
\if!#1!%
\href{https://arxiv.org/abs/math/#2}{\tt math/#2}%
\else%
\href{https://arxiv.org/abs/math.#1/#2}{\tt math.#1/#2}%
\fi}

\title{\boldmath Scalar anomaly cancellation reveals the hidden superalgebraic structure of the quantum chiral $\mathrm{SU}(2/1)$ model of leptons and quarks}

\author{Jean Thierry-Mieg}

\affiliation{NCBI, National Library of Medicine, National Institute of Health, \\
  8600 Rockville Pike, Bethesda MD20894, U.S.A.}

\emailAdd{mieg@ncbi.nlm.nih.gov}

\abstract{At the classical level, the SU(2/1) superalgebra offers a natural description of the elementary particles: leptons and quarks massless states, graded by their chirality, fit the smallest irreducible representations of SU(2/1). Our new proposition is to pair the left/right space-time chirality with the superalgebra chirality and to study the model at the one-loop quantum level. If, despite the fact that they are non-Hermitian, we use the odd matrices of SU(2/1) to minimally couple an oriented complex Higgs scalar field to the chiral Fermions, novel anomalies occur. They affect the scalar propagators and vertices. However, these undesired new terms cancel out, together with the Adler-Bell-Jackiw vector anomalies, because the quarks compensate the leptons. The unexpected and striking consequence is that the scalar propagator must be  normalized using the antisymmetric super-Killing metric and the scalar-vector vertex must use the  symmetric $d\_{aij}$ structure constants of the superalgebra. Despite this extraordinary structure, the resulting Lagrangian is actually Hermitian.}

\pgfkeys{/tikzfeynman/warn luatex=false}
\makeatletter\def\tikzfeynman@luatex@required@path{\relax}\makeatother

\begin{document}
\maketitle
\flushbottom

\section{Background}

The weak interactions are chiral.
All the left helicity states of the leptons and the quarks are
weak $\SU(2)$ doublets, whereas all their right helicity states
are $\SU(2)$ singlets.

This fundamental asymmetry, first recognized by Lee and Yang in 1957, remains a challenge to the
algebraic classification of the elementary particles
because the Lie algebra Yang-Mills multiplets can only describe massless Fermions of
a given chirality, hence cannot unify the
different helicity states of the particles.
Two avenues have been explored. On the one hand, in the grand unified theories, the
anti right-singlets, which are left anti-singlets, are combined
with the left doublets. As particles are coupled to antiparticles,
the baryon number is not conserved and an observable slow decay rate of the proton
is predicted. But this proton decay was not observed in dedicated experiments.
On the other hand, in supersymmetric models,
each known particle must be associated to a new particle: the s-electron, s-quark,
gluino and so on. But the CERN hadron collider has revealed no new physics below 1\,Tev.
Both approaches thus seem incompatible with experiments.
With hindsight, these models did not mark `the end of physics', and the door remains open
to alternative ideas.

In 1979 Ne'eman~\cite{N1} and Fairlie~\cite{F1} proposed to
embed $\SU(2)\U(1)$ in the Lie-Kac superalgebra $\SU(2/1)$.
Their paradigm is to use the chirality $\chi$  as the fundamental $Z(2)$
grading of the superalgebra~\cite{Kac1},
allowing the unification of left and right Fermion states in
 graded multiplets.
The germ of this idea can be traced back to
the original presentation
of the electroweak unification by Weinberg in 1967~\cite{W67}
where he noticed that since there is no massless particle coupled
to the electron number, the $\U(1)$ gauge field
must be proportional to the electronic
hypercharge $Y=N_R + N_L/2$  which has the same trace
over the left and the right leptons
\BE  \label{eq1}
Tr_L(Y) - Tr_R(Y) =  Tr(\chi\;Y) = STr(Y) = 0\,,
\EE
precisely the condition
allowing to embed $\SU(2)\U(1)$ inside $\SU(2/1)$.

The cancellation of the Adler-Bell-Jackiw anomaly~\cite {Adler, BJ},
\BE  \label{eq2}
  C_{abc} = STr (\lX_a, \{\lX_b,\lX_c\}_+) = 0
\EE
gives another indication as it involves a supertrace and an anticommutator
and corresponds to the even part of the cubic super-Casimir tensor of $\SU(2/1)$ (appendix~\ref{A1}  equation~\eqref{eq29} and appendix~\ref{A9}  equation~\eqref{eq65}).

At the static classification level, the $\SU(2/1)$ model is successful.
The leptons~\cite{N1,F1}, the quarks~\cite{DJ,NTM1}, and their antiparticles are naturally
described (appendix~\ref{A2}  and~\ref{A4}) by the lowest dimensional $\SU(2/1)$ irreducible representations~\cite{Kac1,SNR},
unifying in the same chiral multiplets the left and the right massless states.
In addition, contrary to Lie algebras, superalgebras admit finite dimensional
indecomposable representations~\cite{Marcu,GQS,Yucai}, which in the case of $\SU(2/1)$
can regroup at most three generations of quarks (appendix~\ref{A8}  \mbox{and~\cite{CQ0,CQ2,HS,HPS}}).

In other words, $\SU(2/1)$ offers an ideal algebraic classification of
all the existing fundamental Fermions: unlike GUTs $\SU(2/1)$ does not predict
proton decay, unlike SUSY $\SU(2/1)$ does not require the
existence of new Fermions,  yet $\SU(2/1)$ is the only algebraic model which
naturally predicts the number of generations of leptons and quarks.

The symmetry breaking pattern of the adjoint representation is also satisfactory.
Following Yang and Mills, the $\SU(2)\U(1)$ even generators are gauged by the $W^{\pm}$, the $Z^0$ and the photon.
We then postulate that scalar fields correspond to the odd generators.
So  if they acquire a non-zero vacuum expectation value $v$, then $v$ selects one
of the odd directions. But since in a superalgebra the odd generators close
by anticommutation on the even generators, the square $\gamma = \{v,v\}_+$ of the
vacuum automatically corresponds to an even generator that we can identify as the photon.
The super-Jacobi identity (appendix~\ref{A1}, equation~\eqref{eq26}) then implies that the photon commutes with $v$:
\BE
 \label{eq3}   [\gamma, v] = [\{v,v\}_+,v]_- = 0
\EE
and remains massless~\cite {NTM1,CQ1,NSF}.
In 1995, using $\SU(2/1)$ in a qualitative way, Hwang, Lee and Ne'eman~\cite{HLN}
correctly predicted the mass of the Higgs to be $130 \pm 6 $\,Gev,
seventeen years before the experimental observation at 125\,Gev.

The difficulty in the $\SU(2/1)$ model is to extend the Lie algebra Yang-Mills formalism to the
more complex case of a superalgebra. Our proposition is to bypass
the construction of the classical theory and directly study
the Fermions quantum one-loop counterterms which can be computed just
from the assumption that the Bosons are coupled to the Fermions
according to the matrices $\lX$ of the relevant linear representation of the
$\SU(2/1)$ superalgebra.
This has never been attempted, probably because
the quarks odd matrices, listed in appendix~\ref{A6}, are not
Hermitian. Analyzing the scalar propagator and vertices counter-terms,
we show below that because our new scalar-Fermions couplings are non Hermitian and chiral,
the counterterms contain a regular part and an anomaly.
Our surprising discovery is that, exactly like in the Adler-Bell-Jackiw triangle diagrams,
the sum of the lepton and quark contributions~\cite{BIM} cancels out these new scalar anomalies,
whereas, as shown in equation (13), the regular counterterms induce a scalar Lagrangian
\BQA
\LAG_{\Phi} &= - g_{ij}\;D_{\mu}\overline\Phi^i\;D^{\mu}\Phi^j\,.
 \label{eq4}\\
  g_{ij} &= \frac {1}{2} STr (\lX_i\lX_j)\,, \quad D_{\mu}\Phi_i = \partial_{\mu}\Phi_i + d_{aij}A^a_{\mu} \Phi^j\,.
\EQA
exactly as expected of a minimally coupled superalgebra,
where the normalization of the scalar propagator is proportional to the $g_{ij}$ super-Killing metric
and the regular vector-scalar counterterm is proportional to the $d_{aij}$ symmetric structure constants of $\SU(2/1)$.
Despite this unusual structure, the theory is
unitary because a linear change of variables given in equation~\eqref{eq17}
leads back to a classic Lie algebra Hermitian Lagrangian.

In other words, we are not constructing a locally supersymmetric version of the
standard model, but we reveal, at the quantum dynamical level,
the existence inside the model of several new hidden layers of $\SU(2/1)$ superalgebraic
structures.

In the following sections, we present our new results. But since we realize that
the $\SU(2/1)$ model is not well known, we recall in the appendices the definition of
a chiral superalgebra, the construction of the leptons and quarks
$\SU(2/1)$ irreducible or indecomposable representations,
and the principal steps in the calculation of the Adler-Bell-Jackiw vector anomaly.

\section{The chiral scalar-Fermion minimal coupling}

Let us assume the existence of an oriented complex scalar field $\overline{\Phi}\;\Phi$ coupled to the chiral
Fermions $\overline{\psi}\psi$ via the odd generators $\lX_i$ of the superalgebra
The scalars are oriented: they transport left spin states, they are emitted by left $\psi_L$
Fermions
(which then become right) and
absorbed by right $\psi_R$ Fermions (which then become left) according to the Feynman
diagrams:

\newpage

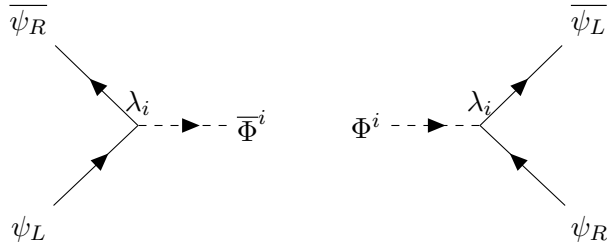
\begin{center}
\begin{tikzpicture}
\begin{feynman}
\vertex (a) {$\PhiB^i$};
\vertex [left = of a, label=$\lX_i$] (x);
\vertex [below left=of x] (b){$\psi_L$};
\vertex [above left=of x] (c){$\overline{\psi_R}$};
\diagram* {
  (a) -- [anti charged scalar] (x),
  (x) --  [anti fermion](b),
  (x) --  [fermion](c),
};
\vertex  [right = of a] (a2) {$\Phi^i$};
\vertex [right = of a2, label=$\lX_i$] (x2);
\vertex [below right=of x2] (b2){$\psi_R$};
\vertex [above right=of x2] (c2){$\overline{\psi_L}$};
\diagram* {
  (a2) -- [charged scalar] (x2),
  (x2) --  [anti fermion](b2),
  (x2) --  [fermion](c2),
};
\end{feynman}
\end{tikzpicture}
\end{center}

To preserve CP invariance, we need to multiply the odd matrices $\lX_i$ by a chiral projector
\BQ  \label{eq5}
 \epsilon_L = \frac{1}{2}(1 + \chi)
\,,\quad
 \epsilon_R = \frac{1}{2}(1 - \chi)
\,.
\EQ
The chirality operator $\chi$, which acts on the algebra charges and defines the supertrace (appendix~\ref{A1}, equation~\eqref{eq23}),
is correlated with the Lorentz chirality operator $\gamma_5$, which acts on the spin indices.
$\Phi$ is absorbed by an $\SU(2)$ singlet right-spinor $\psi_R = 1/4(1-\chi)(1-\gamma_5)\psi_R$,
emitting an $\SU(2)$ doublet left-spinor $\psi_L = 1/4(1+\chi)(1+\gamma_5)\psi_L$.
This correlation explains why the weak interactions break $C$ and $P$ but conserve $CP$.
There is no equivalent relation in the Yang-Mills-Lie algebra framework because the charge chirality $\chi$
is specific of superlagebras. The Fermion-scalar interaction terms of the Lagrangian read:
\BE  \label{eq6}
\LAG_{\psi\Phi} = \overline{(\psi_L)}_R\,\Phi^i\epsilon_L\lX_i\,\psi_R + \overline{(\psi_R)}_L\,\overline\Phi^i\epsilon_R\lX_i\,\psi_L
\,.
\EE

For the moment, we do not specify the Lagrangian of the $\Phi$ scalars.
The idea is to deduce the nature of the propagator of the
scalars and their interactions with the vector fields from
the calculation of the Fermion loops. Consider first the propagator:

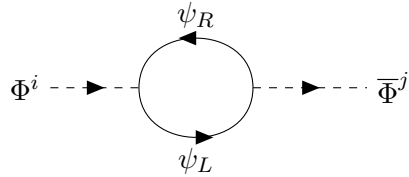
\begin{center}
\begin{tikzpicture}
\begin{feynman}
\vertex (a) {$\Phi^i$};
\vertex [right=of a] (b);
\vertex [right=of b] (c);
\vertex [right=of c] (d){$\PhiB^j$};
\diagram* {
  (a) -- [charged scalar] (b),
  (b) --  [anti fermion, half left, edge label =$\psi_R$ ](c),
  (c) -- [anti fermion, half left,, edge label =$\psi_L$ ] (b),
  (c) -- [charged scalar] (d),
};
\end{feynman}
\end{tikzpicture}
\end{center}

This counterterm
is, as it should, proportional to the inverse square of the momentum $p$ of the
propagating scalar ($1/p^2$), but the trace over the odd matrices is
chiral. We get
\BE  \label{eq7}
   Tr (\epsilon_L \;\lambda_i\;\lambda_j) = \frac {1}{2} \; STr (\lambda_i\;\lambda_j) + \frac {1}{2} \; Tr (\lambda_i\;\lambda_j)
\EE
We like the first term which gives the odd part of the super-Killing metric of the superalgebra.
The second term gives the `would be' symmetric metric of a Lie algebra,
but is not an invariant of a superalgebra. Generalizing the Adler-Bell-Jackiw condition~\eqref{eq2}, we
call it anomalous and request that the combined contribution of all chiral Fermions
vanishes:
\BQ  \label{eq8}
Tr (\lambda_i\;\lambda_j) = 0\,.
\EQ

We now consider the scalar-scalar-vector triangle diagram. There are only two diagrams corresponding to the two possible orientations of the Fermion loop, versus the four diagrams shown in appendix~\ref{A9}  in the case of the vector anomaly. Since
the Fermion loop absorbs $\Phi^i$ and emits $\PhiB^j$,
the orientation of the loop imposes the chirality.

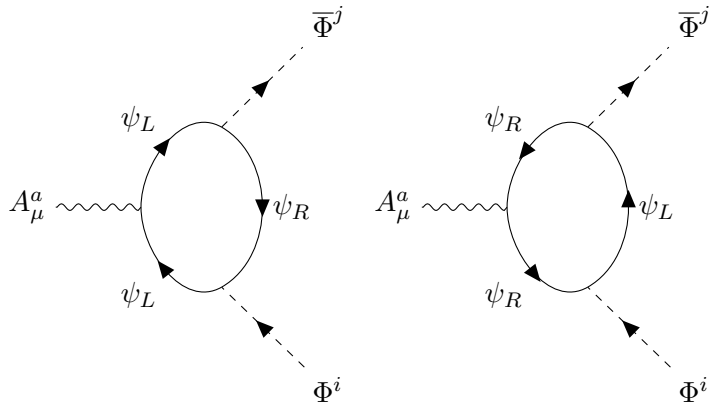
\begin{center}
\begin{tikzpicture}
\begin{feynman}
  \vertex (a1) {$A^a_{\mu}$};
\vertex [right= of a1] (a);
\vertex [below right=of a] (b);
\vertex [above right=of a] (c);
  \vertex [below right=of b](b1) {$\Phi^i$};
  \vertex [above right=of c](c1) {$\PhiB^j$};
\diagram* {
  (a1) -- [photon] (a),
  (b1) -- [charged scalar] (b),
  (c) -- [charged scalar] (c1),
  (a) --  [fermion, in=150,out=90, edge label =$\psi_L$ ](c),
  (c) -- [fermion,  in=30,out=-30, edge label =$\psi_R$ ] (b),
  (b) -- [fermion, in=270,out=210, edge label =$\psi_L$ ] (a),
};
\end{feynman}
\end{tikzpicture}
\begin{tikzpicture}
\begin{feynman}
  \vertex (a1) {$A^a_{\mu}$};
\vertex [right= of a1] (a);
\vertex [below right=of a] (b);
\vertex [above right=of a] (c);
  \vertex [below right=of b](b1) {$\Phi^i$};
  \vertex [above right=of c](c1) {$\PhiB^j$};
\diagram* {
  (a1) -- [photon] (a),
  (b1) -- [charged scalar] (b),
  (c) -- [charged scalar] (c1),
  (a) --  [anti fermion, in=150,out=90, edge label =$\psi_R$ ](c),
  (c) -- [anti fermion, in=30,out=-30, edge label =$\psi_L$ ] (b),
  (b) -- [anti fermion, in=270,out=210, edge label =$\psi_R$ ] (a),
};
\end{feynman}
\end{tikzpicture}
\end{center}

In one orientation,
the vector $A^a_{\mu}$ touches a left Fermion, in the opposite orientation
it touches a right Fermion, and as recalled in appendix~\ref{A9}  for the Adler-Bell-Jackiw triangle diagram,
the orientation governs the overall sign of the diagram. Hence we obtain the unusual term
\BE
 \label{eq9}   Tr (\epsilon_L \;\lambda_a\;\lambda_i\;\lambda_j - \epsilon_R \;\lambda_a\;\lambda_j\;\lambda_i) =
      \frac {1}{2} STr (\lambda_a\;\{\lambda_i\;,\lambda_j\}_+) +
       \frac {1}{2} Tr (\lambda_a\;[\lambda_i\;,\lambda_j]_-) \,.
\EE
The first term of~\eqref{eq9} gives, for any representation of the superalgebra, the symmetric structure constants
of the superalgebra (appendix~\ref{A1}, equations~\eqref{eq25} and~\eqref{eq28}):
\BE  \label{eq10}
d_{aij} =  \frac {1}{2} STr (\lambda_a\;\{\lambda_i\;,\lambda_j\}_+)
\EE
The second term of~\eqref{eq9} gives the `would be' antisymmetric constants
\BE  \label{eq11}
f_{aij} = \frac {1}{2} Tr (\lambda_a\;[\lambda_i\;,\lambda_j]_-)
\EE
which are not well defined, because the commutators of the odd matrices do not close on the
even matrices. We call this second term anomalous, and generalizing the Adler-Bell-Jackiw condition~\eqref{eq2} we request that:
\BE  \label{eq12}
Tr (\lambda_a\;[\lambda_i\;,\lambda_j]_-) \ = 0\,.
\EE

Our surprising result is that the three conditions~\eqref{eq2},~\eqref{eq8} and~\eqref{eq12} are met simultaneously
when we apply the experimentally validated Bouchiat-Iliopoulos-Meyer prescription:
3 quarks for every lepton~\cite{BIM}. In other words,
the propagator~\eqref{eq8} and vertex~\eqref{eq12} scalar anomalies vanish, provided the Adler-Bell-Jackiw anomaly~\eqref{eq2} vanishes.
The 3 conditions are verified by direct examination of the quark and lepton
matrices listed in appendix~\ref{A2}  and~\ref{A4}. The three anomalies also vanish if we consider the antileptons and antiquarks
matrices listed in appendix~\ref{A3} and~\ref{A5}.

Therefore, the renormalization rules~\eqref{eq7},~\eqref{eq9} imply that
the Lagrangian of the scalar field
is explicitly supercovariant:
\BE  \label{eq13}
\LAG_{\Phi} = - g_{ij}\; D_{\mu} \overline{\Phi}^i\;D_{\mu}\Phi^j\,,\quad
D_{\mu}\Phi_i = \partial_{\mu}\Phi_i + d_{aij}\;A^a_{\mu}\;\Phi^j\,,
\EE
where $g_{ij}$ is the antisymmetric super-Killing  metric (appendix~\ref{A1}, equation~\eqref{eq27}) and the supercovariant derivative
$D_{\mu}$ produces the $(ij)$ vertex $d_{aij}\;(p+q)_{\mu}$ where
the $d_{aij}$ are the symmetric structure constants of the superalgebra (appendix~\ref{A1}, equation~\eqref{eq25}),
and  p and q are the momenta of the
incoming and outgoing $\Phi$ fields in the orientation of the $\Phi$~lines.

Finally, we consider the $AA\PhiB\Phi$ two-vectors-two-scalars vertex
which gives an additional constraint.

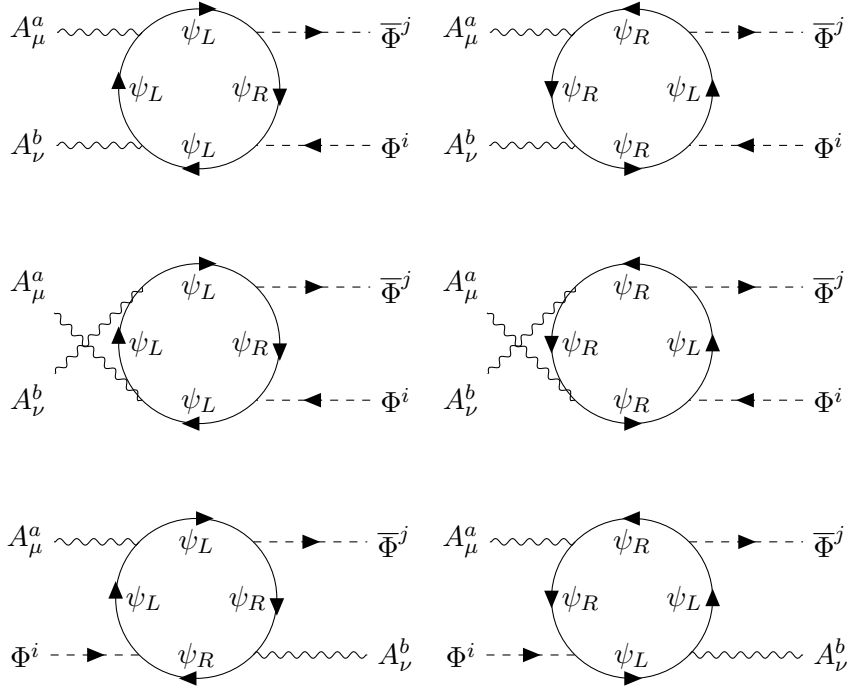
\begin{center}
\begin{tikzpicture}
\begin{feynman}
  \vertex (a1) {$A^a_{\mu}$};
  \vertex [below= of a1] (a2) {$A^b_{\nu}$};
  \vertex [right= of a1] (b1);
  \vertex [right= of a2] (b2);
  \vertex [right= of b2] (b3);
  \vertex [right= of b1] (b4);
  \vertex [right= of b3] (a3) {$\Phi^i$};
  \vertex [right= of b4] (a4) {$\PhiB^j$};
\diagram* {
  (a1) -- [photon] (b1),
  (a2) -- [photon] (b2),
  (a3) -- [charged scalar] (b3),
  (b4) -- [charged scalar] (a4),
  (b1) --  [anti fermion, quarter right, edge label =$\psi_L$ ](b2),
  (b2) --  [anti fermion, quarter right, edge label =$\psi_L$ ](b3),
  (b3) --  [anti fermion, quarter right, edge label =$\psi_R$ ](b4),
  (b4) --  [anti fermion, quarter right, edge label =$\psi_L$ ](b1),
};
\end{feynman}
\end{tikzpicture}
\begin{tikzpicture}
\begin{feynman}
  \vertex (a1) {$A^a_{\mu}$};
  \vertex [below= of a1] (a2) {$A^b_{\nu}$};
  \vertex [right= of a1] (b1);
  \vertex [right= of a2] (b2);
  \vertex [right= of b2] (b3);
  \vertex [right= of b1] (b4);
  \vertex [right= of b3] (a3) {$\Phi^i$};
  \vertex [right= of b4] (a4) {$\PhiB^j$};
\diagram* {
  (a1) -- [photon] (b1),
  (a2) -- [photon] (b2),
  (a3) -- [charged scalar] (b3),
  (b4) -- [charged scalar] (a4),
  (b1) --  [fermion, quarter right, edge label =$\psi_R$ ](b2),
  (b2) --  [fermion, quarter right, edge label =$\psi_R$ ](b3),
  (b3) --  [fermion, quarter right, edge label =$\psi_L$ ](b4),
  (b4) --  [fermion, quarter right, edge label =$\psi_R$ ](b1),
};
\end{feynman}
\end{tikzpicture}
\vskip 1cm
\begin{tikzpicture}
\begin{feynman}
  \vertex (a1) {$A^a_{\mu}$};
  \vertex [below= of a1] (a2) {$A^b_{\nu}$};
  \vertex [right= of a1] (b1);
  \vertex [right= of a2] (b2);
  \vertex [right= of b2] (b3);
  \vertex [right= of b1] (b4);
  \vertex [right= of b3] (a3) {$\Phi^i$};
  \vertex [right= of b4] (a4) {$\PhiB^j$};
\diagram* {
  (a1) -- [photon] (b2),
  (a2) -- [photon] (b1),
  (a3) -- [charged scalar] (b3),
  (b4) -- [charged scalar] (a4),
  (b1) --  [anti fermion, quarter right, edge label =$\psi_L$ ](b2),
  (b2) --  [anti fermion, quarter right, edge label =$\psi_L$ ](b3),
  (b3) --  [anti fermion, quarter right, edge label =$\psi_R$ ](b4),
  (b4) --  [anti fermion, quarter right, edge label =$\psi_L$ ](b1),
};
\end{feynman}
\end{tikzpicture}
\begin{tikzpicture}
\begin{feynman}
  \vertex (a1) {$A^a_{\mu}$};
  \vertex [below= of a1] (a2) {$A^b_{\nu}$};
  \vertex [right= of a1] (b1);
  \vertex [right= of a2] (b2);
  \vertex [right= of b2] (b3);
  \vertex [right= of b1] (b4);
  \vertex [right= of b3] (a3) {$\Phi^i$};
  \vertex [right= of b4] (a4) {$\PhiB^j$};
\diagram* {
  (a1) -- [photon] (b2),
  (a2) -- [photon] (b1),
  (a3) -- [charged scalar] (b3),
  (b4) -- [charged scalar] (a4),
  (b1) --  [fermion, quarter right, edge label =$\psi_R$ ](b2),
  (b2) --  [fermion, quarter right, edge label =$\psi_R$ ](b3),
  (b3) --  [fermion, quarter right, edge label =$\psi_L$ ](b4),
  (b4) --  [fermion, quarter right, edge label =$\psi_R$ ](b1),
};
\end{feynman}
\end{tikzpicture}
\vskip 1cm

\begin{tikzpicture}
\begin{feynman}
  \vertex (a1) {$A^a_{\mu}$};
  \vertex [below= of a1] (a2) {$\Phi^i$};
  \vertex [right= of a1] (b1);
  \vertex [right= of a2] (b2);
  \vertex [right= of b2] (b3);
  \vertex [right= of b1] (b4);
  \vertex [right= of b3] (a3) {$A^b_{\nu}$};
  \vertex [right= of b4] (a4) {$\PhiB^j$};

\diagram* {
  (a1) -- [photon] (b1),
  (a2) -- [charged scalar] (b2),
  (a3) -- [photon] (b3),
  (b4) -- [charged scalar] (a4),
  (b1) --  [anti fermion, quarter right, edge label =$\psi_L$ ](b2),
  (b2) --  [anti fermion, quarter right, edge label =$\psi_R$ ](b3),
  (b3) --  [anti fermion, quarter right, edge label =$\psi_R$ ](b4),
  (b4) --  [anti fermion, quarter right, edge label =$\psi_L$ ](b1),

};
\end{feynman}
\end{tikzpicture}
\begin{tikzpicture}
\begin{feynman}
  \vertex (a1) {$A^a_{\mu}$};
  \vertex [below= of a1] (a2) {$\Phi^i$};
  \vertex [right= of a1] (b1);
  \vertex [right= of a2] (b2);
  \vertex [right= of b2] (b3);
  \vertex [right= of b1] (b4);
  \vertex [right= of b3] (a3) {$A^b_{\nu}$};
  \vertex [right= of b4] (a4) {$\PhiB^j$};

\diagram* {
  (a1) -- [photon] (b1),
  (a2) -- [charged scalar] (b2),
  (a3) -- [photon] (b3),
  (b4) -- [charged scalar] (a4),
  (b1) --  [fermion, quarter right, edge label =$\psi_R$ ](b2),
  (b2) --  [fermion, quarter right, edge label =$\psi_L$ ](b3),
  (b3) --  [fermion, quarter right, edge label =$\psi_L$ ](b4),
  (b4) --  [fermion, quarter right, edge label =$\psi_R$ ](b1),

};
\end{feynman}
\end{tikzpicture}
\end{center}

The diagrams are symmetrized in $(a\mu,b\nu)$ but not in $(ij)$ since
$\Phi$ and $\PhiB$ are distinct.
Carefully computing the trace of six
$\SG$ matrices (appendix~\ref{A9}, equations~\eqref{eq63}--\eqref{eq64}), we find that the counterterm is proportional to
\BE  \label{eq14}
Tr ((\lX_a\lX_b+\lX_b\lX_a)(\epsilon_L\lX_i\lX_j + \epsilon_R\lX_j\lX_i) - 2 ( \epsilon_L \lX_a \lX_i \lX_b \lX_j + \epsilon_R \lX_a \lX_j \lX_b \lX_i)).
\EE
This trace can be decomposed into the sum of two terms
\BE  \label{eq15}
g^{ij} (d_{aik} d_{bjl} + d_{bik} d_{ajl}) + \Delta(\rho) \delta_{ij} (f^i_{ak} f^j_{bl} + f^i_{bk} f^j_{al})\,.
\EE
We like the first term of this equation. It is proportional to $(d_{a..})^2$ which is characteristic of a superalgebra.
It is representation independent.
It matches the term $g^{\mu\nu}A^a_{\mu}A^b_{\nu}\PhiB^k\Phi^l$ $g^{ij} (d_{aik} d_{bjl}+ d_{bik} d_{ajl})$
present in the classical Lagrangian~\eqref{eq13}. Therefore, it can be absorbed by a renormalization of the
coupling constant $\gcpl^2$. The relative renormalization of $\gcpl$ in the
$\PhiB\Phi$, $\gcpl\,A\PhiB\Phi$
and $\gcpl^2\,AA\PhiB\Phi$ diagrams is correct because the integrals
over the loop-momenta are the same as in the standard Yang-Mill-scalar theory, only the
group traces are new. The second term of~\eqref{eq15} is proportional to $(f_{a..})^2$ which is characteristic of a
'would be' Lie algebra. Its normalization $\Delta(\rho)$ depends on the representation.
We call this term anomalous and verified numerically, with a simple C-program,
that the combined quark and lepton contributions again cancel out thanks to the BIM mechanism~\cite{BIM}
\BE  \label{eq16}
\Delta (\text{leptons}) \neq 0\,,\quad \Delta(\text{leptons}) + 3 \Delta(\text{quarks}) = 0\,.
\EE
In conclusion, the $AA\PhiB\Phi$ term is renormalizable, establishing a new scalar generalization of
the Ward, Takahashi, Slavnov, Taylor identity to the case of the $\SU(2/1)$ superalgebra.

As shown at the end of appendix~\ref{A5}, any combination of leptons and quark-like
representations such that the total sum of the hypercharges of the left doublets vanishes is anomaly free.
We already discussed the standard model assignment, one electron of hypercharge $-1$ and 3 colors of quarks
of hypercharge $1/3$, but we
could also consider the $\OSp(2/1)$ neutral representation of Minahan, Ramon and Warner (appendix~\ref{A5} and~\cite{MRW}),
or one quark doublet of hypercharge $2/3$ and two of hypercharge $-1/3$, and so on.
We leave as an open problem the general classification of all the
chiral representations of the simple superalgebras satisfying the
four equations~\eqref{eq2},~\eqref{eq8},~\eqref{eq12},~\eqref{eq16}
and conjecture that these anomalies play a role in the exponentiation of the superalgebra
into a supergroup.

These results are unexpected and were not anticipated in the
$\SU(2/1)$ literature. It was known since the early eighties
that the quantum numbers of quarks and leptons corresponded to the $\SU(2/1)$
irreducible representations, but there was no sign that the
superalgebra metric and the $d_{aij}$ superstructure constants could play a role
in the dynamics of the theory.

A vertex proportional to the $d_{aij}$ symmetric structure constant is actually a necessity
in a superalgebraic theory. Consider the renormalization of the vector-Fermion vertex
where the vector $A^a_{\mu}$ emits a pair $\PhiB^i\Phi^j$ via a vertex $h_{aij}$
with unknown $(ij)$ symmetry. The 2 scalars then hit the Fermion generating a
matrix product $h_{aji}\,\lambda_i\lambda_j$:

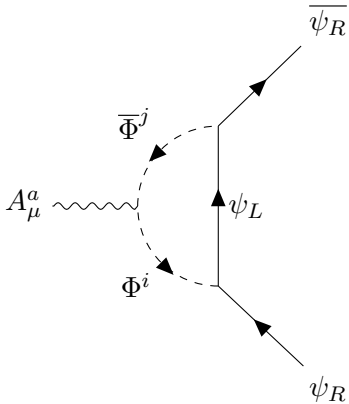
\begin{center}
\begin{tikzpicture}
\begin{feynman}
  \vertex (a1) {$A^a_{\mu}$};
\vertex [right= of a1] (a);
\vertex [below right=of a] (b);
\vertex [above right=of a] (c);
  \vertex [below right=of b](b1) {$\psi_R$};
  \vertex [above right=of c](c1) {$\overline{\psi_R}$};
\diagram* {
  (a1) -- [photon] (a),
  (b1) -- [fermion] (b),
  (c) -- [fermion] (c1),
  (a) --  [anti charged scalar, quarter left, edge label =$\PhiB^j$ ](c),
  (c) -- [anti fermion, edge label =$\psi_L$ ] (b),
  (b) -- [anti charged scalar, quarter left, edge label =$\Phi^i$ ] (a),
};
\end{feynman}
\end{tikzpicture}
\end{center}

In the classic Yang-Mills case,
the vector scalar vertex $f_{aij}$ is antisymmetric in $(ij)$ generating the commutator
$f_{aij}\,[\lambda_i\lambda_j]$
which closes on $\lambda_a$. But in a superalgebra, we
need an anticommutator, so $h_{aij}$ has to be symmetric in $(ij)$
and coincides with $d_{aij}$.
With Yuval Ne'eman, we were already hoping to
solve this difficulty in 1982 by representing the odd generators
using higher forms~\cite{NTM2, Quillen}, but that method did not
produce the desired effect.
This is why, after all these years, I am so pleased
and so surprised by the new concept of the scalar anomaly cancellation
presented here.
The solution  lies beyond the
analysis of the abstract superalgebra structure and even
beyond the analysis of its irreducible representation.
It comes from the conspiracy of quarks and leptons.
Separately, they each generate an anomaly, yet together they produce the
desired symmetric vertex.

\section{Rediagonalization to an explicitly Hermitian Lagrangian}

From the analysis of the scalar anomalies, we found a very unusual
structure for the covariant propagator of the $\Phi$ scalars~\eqref{eq13}.
It involves the antisymmetric super-Killing metric $g_{ij}$
and a $d_{aij}$  symmetric structure constant in the
definition of the covariant derivative.
In addition, the scalar-Fermion vertex involves
the non Hermitian odd matrices $\lX_i$.

To show that this theory is nevertheless consistent,
we define new scalar fields $H$ and $K$ by the
linear equations
\BQA
\Phi^4 &= \frac{1}{2}\;(H^4 - i H^5 + K^5 - i K^4)
\,,\\
\Phi^5 &= \frac{1}{2}\;(H^5 + i H^4 - K^4 - i K^5)
\,,\\
\overline{\Phi}^4 &= \frac{1}{2}\;(H^4 + i H^5 - K^5 - i K^4)
\,,\\
\overline{\Phi}^5 &= \frac{1}{2}\;(-i H^4 + H^5 + K^4 -i K^5)
\,,\\
\Phi^6 &= \frac{1}{2}\;(H^6 - i H^7 + K^7 - i K^6)
\,,\\
\Phi^7 &= \frac{1}{2}\;(H^7 + i H^6 - K^6 - i K^7)
\,,\\
\overline{\Phi}^6 &= \frac{1}{2}\;(H^6 + i H^7 - K^7 - i K^6)
\,,\\
\overline{\Phi}^7 &= \frac{1}{2}\;(-i H^6 + H^7 + K^6 -i K^7)
\,,  \label{eq17}
\EQA
By substitution, we find that the super-Killing antisymmetric propagator~\eqref{eq13} of the
oriented $\Phi$ fields can be rewritten as a standard
positive-defined diagonal propagator
for the $H$ and $K$ scalar fields
\BE  \label{eq18}
g_{ij}\;\partial^{\mu}\overline{\Phi}^i\,\partial_{\mu}\Phi^j
= - \frac{1}{2} \delta_{ij}\;(\partial^{\mu}H^i\,\partial_{\mu}H^j+\partial^{\mu}K^i\,\partial_{\mu}K^j)\,.
\EE
The couplings of the scalars to the Yang-Mills vectors also become
standard. By substitution, we find that the unusual superalgebraic
$d_{aij}$ symmetric vertex~\eqref{eq13} reverts to $\SU(2)\U(1)$ minimal couplings
\BE  \label{eq19}
d_{aij} \;A^a_{\mu}\;(\partial^{\mu}\overline{\Phi}^i\; \Phi^j
+ \Phi^i\;\partial^{\mu}\overline{\Phi}^j)
= f_{aij} \;A^a_{\mu}\;(H^i\overleftrightarrow{\partial}^{\mu}H^j + K^i\overleftrightarrow{\partial}^{\mu}K^j)
\EE
\sloppy{Even more surprising, the non Hermitian couplings of $\Phi$ scalars  to the chiral
Fermions~\eqref{eq2},~\eqref{eq5} mutate into Hermitian couplings of the $H$ and $K$ scalars}
\BE  \label{eq20}
\epsilon_R\overline{\Phi}^i\lX_i + \epsilon_L\Phi^i\lX_i =
H^i \mu^-_i + K^i \mu^+_i
\EE
where the $\mu^-_i$ matrices correspond to the Hermitian part of the
$\lX_i$ matrices, and therefore interact only with the doublets and the negatively
charged right Fermions singlets, and the $\mu^+_i$ matrices correspond to the
anti-Hermitian part of the $\lX_i$ matrices, and therefore interact
only with the doublets and the positively charged right Fermions singlets.
In the lepton representation (appendix~\ref{A2})
\BE  \label{eq21}
\mu^-_i = \lX_i\,,\quad \mu^+_i = 0\,,\qquad i= 4,5,6,7
\EE
In the quark representation (appendix~\ref{A4})
\BE  \label{eq22}
\mu^-_6 = \frac {1}{\st}  \begin{pmatrix}
 0 & 0 & 0 & 0 \cr 0 & 0 & 0 & 0 \cr 0 & 0 & 0 & 1
 \cr 0 & 0 & 1 & 0
\end{pmatrix}
\,,\quad
\mu^+_6 = \frac {1}{\st}  \begin{pmatrix}
 0 & -i\sd & 0 & 0 \cr i\sd & 0 & 0 & 0 \cr 0 & 0 & 0 & 0
 \cr 0 & 0 & 0 & 0
\end{pmatrix}
\,.
\EE
The other odd matrices $\mu_4,\mu_5,\mu_7$ follow the same pattern and are given in appendix~\ref{A6}.

The $H$ and $K$ fields have been defined previously by Haussling and Scheck in~\cite{HS,HPS}, but without proper justification.
Noticing that the odd quark matrices $\lX_i$ are non Hermitian, they added
to the natural scalar-Fermion Lagrangian~\eqref{eq6}
its Hermitian conjugate $\LAG^{\dagger}$, in a way double-counting the particles and the antiparticles.
This induced the same Hermitian scalar-Fermion coupling
$Hi\mu_i^- + K^i\mu_i^+$~\eqref{eq20}--\eqref{eq22},
but they could not relate $H$ and $K$ to $\PhiB\Phi$
because they implicitly assumed that the $\PhiB\Phi$ Lagrangian
  has the usual structure $\LAG = \delta_{ij} D^{\mu}\PhiB^i D_{\mu}\Phi^j$
  with $D_{\mu}\Phi^j = \partial_{\mu}\Phi^j + f^j_{ak}\Phi^k$.

In conclusion, using an axiomatic top-down approach, we have
discovered that the `Standard Model' equipped with a
conventional complex Higgs scalar $\SU(2)$ doublet $H + iK$
hides an explicit superalgebraic structure, which is
revealed by rewriting the $H$ and $K$ fields in terms
of the superalgebraic $\overline{\Phi}$ and $\Phi$ fields using
the linear equation~\eqref{eq17}. Furthermore, if we start from the antileptons and antiquarks
representations, we find exactly the same $H$ $K$ Lagrangian.
 These transformations only make sense in
the quantum world and are implied by the analysis of the anomalies
of the one-loop counterterms.

\section{Generation mixing}

The $H$ and $K$ fields bring us back to the study
by Haussling and Scheck of the indecomposable representations of $\SU(2/1)$~\cite{CQ0,HS,HPS}.
Since these representations
can be written as block triangular matrices (appendix~\ref{A7} and~\ref{A8}),
the mixing terms do not contribute to the calculation of matrix traces,
so they do not modify
our calculation of the anomalies~\eqref{eq2},~\eqref{eq8},~\eqref{eq12},~\eqref{eq16}.
Therefore, the indecomposable representations of $\SU(2/1)$
are admissible and lead to the same definition~\eqref{eq17} of the
$H$ and $K$ fields.

These representations provide, inside the $\SU(2/1)$ framework,
a natural understanding of
neutrino oscillations (\cite{HPS} and appendix~\ref{A7}),
and of the existence of at most
three generations of quarks and leptons with their mixing angles (\cite{HS} and appendix~\ref{A8}),
a schema that no other algebraic model explains.
But since the mixing angles do not play a role in the calculations, the anomaly conditions
do not link the leptons mixing angles to the quarks mixing angles.
Note the direct contradiction with~\cite{NSF} which predicts $2^p$ generations.

This property of superalgebras overcomes an early counter argument of
Feynman (private communication, 1979) who noticed that if the mass of the quarks could be explained
by an irreducible symmetry, then the $up$, $charm$ and $top$ quarks would have the same mass. The solution
of this paradox is that the $\SU(2/1)$ superalgebra admits a single indecomposable representation
which describes at once the three generations
explaining why the quarks have unequal masses and how heavier quarks decay into
lighter quarks.

In contrast to the presentation of the Marseille-Mainz group~\cite{CQ2,HS,HPS},
we believe that all these extraordinary results are direct consequences of
the algebraic properties of the $\SU(2/1)$ superalgebra, and are not related in an obvious way
to the non-commutative geometry of Alain Connes~\cite {CL,C1,C2}.

\section{Limitations of the model}

There remains an important problem in the construction of
a fully consistent $\SU(2/1)$ quantum field theory.
Contrary to the vector-Fermion vertex, the scalar-Fermion
vertex is not protected by the Ward identities.
Therefore the strong interactions
contribute to the  renormalization of the scalar-quark vertex
although they do not affect the scalar-lepton vertex.
As a result, the balance between the leptons and the quarks necessary to cancel the
scalar anomalies does not seem to be preserved at the 2-loop level.
An open problem is to see if this is a genuine obstruction, and if so, can the symmetry can be restored,
for instance by incorporating aspects of the non commutative differential geometry
of Connes~\cite{C1}, or the self-dual scalars of Avdeev-Chizhov~\cite{AC,W}, or
$\OSp(4/2)$ Fermion ghosts~\cite{STMV}, or any new idea.

\section{Discussion}

The weak interactions are chiral. Before symmetry breaking, the leptons and quarks
are massless, their left and right helicity states are distinct,
and only the left states couple to the weak $\SU(2)$ interactions.
As understood by Weinberg~\cite{W67} in 1967, there are no charged
massless Fermions, so the total hypercharge $Y$
of the left and right states must be equal: $Tr_L(Y) - Tr_R(Y) =  Tr(\chi\;Y) = STr(Y) = 0$,
allowing to identify the electroweak $\SU(2)\U(1)$ Lie algebra with the even
part of the Kac superalgebra $\SU(2/1)$, graded by chirality~\cite{N1,F1}.

The same conclusion can be derived from the study of Adler-Bell-Jackiw anomaly~\cite {Adler, BJ}.
Applied to the $\U(1)\SU(3)\SU(3)$ quark loop, we learn that $STr(Y) = 0$
for the quarks. Applied to the $\U(1)\SU(2)\SU(2)$ Fermion loop, we learn,
as discovered by Bouchiat Iliopoulos and Meyer~\cite{BIM}, that
the lepton and the quark diagrams are both anomalous, but the lepton
loop is compensated by three quark loops (BIM mechanism).
Furthermore, the Adler Bell-Jackiw anomaly~\eqref{eq2}
is proportional to the even part of the cubic super-Casimir tensor of $\SU(2/1)$ (appendix~\ref{A1}, equation~\eqref{eq29}).

The purpose of our study is to cast the three families of leptons and quarks
into representations of the $\SU(2/1)$ superalgebra and to
associate the Higgs field to the odd generators.
This idea was first proposed independently in 1979 by Ne'eman~\cite{N1} and Fairlie~\cite{F1}
who observed, as shown in appendix~\ref{A2}, that the (2/1) fundamental representation of  $\SU(2/1)$
fits the leptons $(\nu_l,e_L / e_R)$ graded by their chirality.
The model was rapidly extended to the quarks $(u_R / u_L,d_L / d_R)$ by Dondi, Jarvis, Ne'eman and
Thierry-Mieg~\cite{DJ,NTM1}
which, as shown in appendix~\ref{A4}, fit the smallest typical representation of $\SU(2/1)$~\cite{SNR}.
On the lepton side, as shown in appendix~\ref{A5}, $\SU(2/1)$ specifies that if
the charge of the $e^-$ electron is equal
to the charge of the $W^-$ vector Boson, then the $\U(1)$ charge of the right neutrino
vanishes~\cite{NTM1}. Hence the right neutrino should be weakly neutral, an experimentally validated
prediction. It was then discovered in the nineties~\cite{CQ0,CQ2,HS,HPS} that the indecomposable
representations of $\SU(2/1)$ fit the existence and decays of the heavier families.

A main perceived problem of the $\SU(2/1)$ model is that the odd matrices are non Hermitian.
For example, one can choose a base where the electron odd matrices are Hermitian (appendix~\ref{A2}),
but since the square of the matrix $\lX_6$ gives the electric charge,
it follows that in the antielectron representation,
$(\lX_6)^2$ has the opposite sign, hence the odd antielectron
matrices are anti-Hermitian (appendix~\ref{A3}).
Furthermore, the quark and antiquark odd matrices are neither Hermitian
nor anti-Hermitian (appendix~\ref{A4}).
This complexity seemed to prevent any form of minimal coupling.

But here, we report a discovery. If we strictly apply the
$\SU(2/1)$ representation theory and associate the odd generators of $\SU(2/1)$
to an oriented complex doublet of scalar (Higgs) fields coupling the
left and right Fermions, the non Hermitian character
of the odd matrices generates a new set of anomalies.
The one-loop leptons or quarks contributions to the
self diffusion of the vector Bosons~\eqref{eq2}, to the propagator of the scalars~\eqref{eq8},
and to the diffusion of the scalars by the vector Bosons~\eqref{eq12}
are all anomalous. However, the contributions of the leptons
are exactly compensated by those of the three quarks~\cite{BIM}, canceling at the
same time the Adler-Bell-Jackiw vector anomalies~\cite{Adler,BJ} and the new
scalar anomalies discovered here.
It follows that the propagator of the complex scalars is
given by the odd part of the super-Killing metric of $\SU(2/1)$
and that the $A_{\mu}^a\;\Phi^i\;\Phi^j$ coupling is given by
the $d_{aij}$ symmetric structure constant characteristic of a superalgebra~\eqref{eq13}.
We also establish a superalgebraic scalar Ward identity~\eqref{eq16}
linking the renormalization of the $\PhiB\Phi$ propagator,
$A\PhiB\Phi$ triangle diagram and $AA\PhiB\Phi$ square diagram,
as another new consequence of the BIM mechanism.  All calculations were
done manually and verified
numerically using a simple C-language program.
A linear change of variables~\eqref{eq17} then transforms back this unusual Lagrangian
to a classic model with a pair of scalars $H$ and $K$ respectively
coupled to the up and down right quark states, $u_R$ and $d_R$,
via Hermitian matrices~\eqref{eq21}--\eqref{eq22}, without breaking the algebraic structure
by artificially adding the Hermitian conjugated Lagrangian as was
necessary in~\cite {CQ0,CQ1,CQ2,HS,HPS}.

Although $\SU(2/1)$ is a superalgebra, the present
construction respects the statistics of the particles: the Yang-Mills vectors and the scalars are commuting Bosons,
the leptons and quarks are spin-half anticommuting Fermions,
and all interactions are Lorentz covariant. Rather than changing Bosons into Fermions
like the Wess-Zumino space-time supersymmetry, the $\SU(2/1)$ internal supersymmetry
exchanges the chirality of the Fermions without changing their statistics.
Furthermore, the pairing~\eqref{eq5} of the left/right space-time chirality $\gamma_5$,  with the charge chirality $\chi$
which defines the supertrace of the superalgebra,
provides an algebraic explanation
of the $CP$ structure of the weak interaction which is lacking
in the classic Yang-Mills Lie algebra formalism.

In conclusion, we recall that the $\SU(2)\U(1)$ standard model of the
electroweak interactions contains a hidden chiral $\SU(2/1)$
superalgebraic structure~\cite{N1,F1} which explains
the quantum numbers of the quarks~\cite{DJ,NTM1} using non-Hermitian matrices.
The necessary cancellation of the resulting scalar anomalies dictates the
structure of the scalar Lagrangian, and we have for the first time
established a new kind of minimal coupling of a chiral superalgebra
where the Hermitian Lie subalgebra matrices define as usual the
emission/absorption of the Yang-Mills vector Bosons by the Fermions,
and where the non-Hermitian odd generators define the chirality flipping
absorption/emission of an oriented scalar Higgs field by the chiral Fermions.
In this framework, the super-Killing metric and the $d_{aij}$ superstructure
constants of $\SU(2/1)$ define the propagator and vector diffusion
of the chirality aware Higgs scalars, which naturally complement the Yang-Mills field
in the intrinsic-geometrical definition of the Lie superalgebra chiral connection~\cite{TM20}.

\acknowledgments
This research was supported by the Intramural Research Program of the National Library of Medicine, National Institute of Health.
We are grateful to Danielle Thierry-Mieg for clarifying the presentation.

\appendix

\section{Definition of a chiral superalgebra} \label{A1}

Let us define, using the notations of~\cite{TM20},  a chiral-superalgebra as a finite dimensional
basic classical Lie-Kac superalgebra~\cite{Kac1}, graded by chirality.
For example, we could take a superalgebra of type $\SU(m/n)$,
or $\OSp(m/2n)$ or a product of Lie algebras and superalgebras
like the $\SU(2/1)$ superalgebra of the standard model.

The superalgebra acts on a finite dimensional space
of massless Fermion states graded by their helicity.
The chirality matrix $\chi$ is diagonal, with eigenvalue $1$ on the
left Fermions and $-1$ on the right Fermions. It defines the supertrace
\BE  \label{eq23}
STr(\ldots ) = Tr (\chi\;\ldots )
\EE
Each generator is represented by
a finite dimensional matrix of complex numbers (we do not need anticommuting Grassman numbers).
The even generators are denoted $\lX_a$ and the odd generators $\lX_i$.
$\chi$ commutes with the $\lX_a$ and anticommutes with the $\lX_i$
\BE  \label{eq24}
 [\chi,\;\lX_a]_- = \{\chi,\;\lX_i\}_+ = 0
\EE
The $\lX$ matrices close under (anti)-commutation
\BE  \label{eq25}
  [\lX_a,\;\lX_b]_- = f^c_{ab} \;\lX_c
\,,\quad
  [\lX_a,\;\lX_i]_- = f^j_{ai} \;\lX_j
\,,\quad
  \{\lX_i,\;\lX_i\}_+ = d^a_{ij} \;\lX_a
\,,
\EE
and satisfy the super-Jacobi relation with 3 cyclic permuted terms:
\BE  \label{eq26}
   (-1)^{AC} \{\lX_A, \{\lX_B,\;\lX_C]] +
   (-1)^{BA} \{\lX_B, \{\lX_C,\;\lX_A]] +
   (-1)^{CB} \{\lX_C, \{\lX_A,\;\lX_B]]  = 0\,.
\EE
The quadratic Casimir tensor $(g_{ab},g_{ij})$, also called the super-Killing metric, is defined as
\BQA   \label{eq27}
  g_{ab} &= \frac{1}{2} STr (\lX_a\lX_b)\,,
\\
  g_{ij} &= \frac{1}{2} STr (\lX_i\lX_j)\,.
\EQA
The even part $g_{ab}$ of the metric is as usual symmetric,
but because the odd generators anticommute~\eqref{eq24} with
the chirality hidden in the supertrace~\eqref{eq23},
its odd part $g_{ij}$ is antisymmetric.
The structure constants can be recovered from the
supertrace of a product of 3 matrices
\BQA  \label{eq28}
f_{abc} &= g_{ae}\,f^e_{bc} = \frac{1}{2} STr (\lX_a\,[\lX_b,\lX_c]_-)\,,
\\
d_{aij} &= g_{ae}\,d^e_{ij} = \frac{1}{2} STr (\lX_a\,\{\lX_i,\lX_j\}_+)\,,
\EQA
The cubic Casimir tensor is defined as
\BQA  \label{eq29}
  C_{abc} &= \frac{1}{2} STr (\lX_a\;\{\lX_b,\;\lX_c\}_+)\,,
\\
  C_{aij} &= \frac{1}{2} STr (\lX_a\;[\lX_i,\;\lX_j]_-)\,.
\EQA
The Casimirs use the `wrong' type of commutator,
otherwise, using equation~\eqref{eq25}, they could be simplified.
We have $g_{ai} = C_{abi} = C_{ijk} = 0$ since the diagonal elements of
the product of an odd number of odd matrices necessarily vanish.
Using these tensors, we can
construct the super-Casimir operators
\BE  \label{eq30}
 K_2 = g^{AB}\;\lX_A\lX_B\,,\quad K_3 = C^{ABC}\;\lX_A\lX_B\lX_C\,,
\EE
where the upper index metric $g^{AB}$ is the inverse of the lower metric $g_{AB}$,
summation over the repeated indices is implied and ranges over
even and odd values $A,B = a,b,\ldots ,i,j\ldots $, and the indices of $C^{ABC}$
are raised using $g^{AB}$.
The Casimir operators $K_2$ and $K_3$ commute with all the generators
of the superalgebra. In an
irreducible representation, they are represented by a multiple of the
identity matrix. In $\SU(2/1)$, which has rank 2, they form a basis of
its enveloping superalgebra.

\section{The SU(2/1) lepton representation}\label{A2}

Consider the left neutrino and the left and right electron
states collectively called the leptons $(\nu_L,\;e_L\;/\;e_R)$.
Their experimentally observed chirality and weak hyper-charge
are given by the diagonal matrices
\BQ  \label{eq31}
\chi = \begin{pmatrix}
 1 & 0 & 0 \cr 0 & 1 & 0 \cr 0 & 0 & -1
\end{pmatrix}
,\quad
\lX_0 =   \begin{pmatrix}
 -1 & 0 & 0 \cr 0 & -1 & 0 \cr 0 & 0 & -2
\end{pmatrix}
.
\EQ
With respect to the chiral $Z_2$ grading, the supertrace of $\lX_0$ vanishes:
\BQ  \label{eq32}
 STr (\lX_0) = Tr (\chi\;\lX_0) = 0.
\EQ
This is the first indication that the electroweak interactions could be described by a superalgebra.
In the same coordinates, the $\SU(2)$ weak charges are given by the matrices
\BQ   \label{eq33}
\lX_1 =   \begin{pmatrix}
 0 & 1 & 0 \cr 1 & 0 & 0 \cr 0 & 0 & 0
\end{pmatrix}
,\quad
\lX_2 =   \begin{pmatrix}
 0 & -i & 0 \cr i & 0 & 0 \cr 0 & 0 & 0
\end{pmatrix}
,\quad
\lX_3 =   \begin{pmatrix}
 1 & 0 & 0 \cr 0 & -1 & 0 \cr 0 & 0 & 0
\end{pmatrix} .
\EQ
The four $\lX_a$ matrices ($a=0,1,2,3)$ represent the Lie algebra $\SU(2).\U(1)$.
They close under commutation
\BQ  \label{eq34}
   [\lX_b,\;\lX_c]_- = f^a_{bc}\;\lX_a\qquad a,b,c = 0,1,2,3
\EQ
and the only non zero structure constants are $f^1_{23} = f^2_{31} =  f^3_{12} = 2i$.

Let us now add in the picture the four Hermitian matrices
\BQ  \label{eq35}
\lX_4 =   \begin{pmatrix}
 0 & 0 & 1 \cr 0 & 0 & 0 \cr 1 & 0 & 0
\end{pmatrix}
\quad
\lX_5 =   \begin{pmatrix}
 0 & 0 & -i \cr 0 & 0 & 0 \cr i & 0 & 0
\end{pmatrix}
\quad
\lX_6 =   \begin{pmatrix}
 0 & 0 & 0 \cr 0 & 0 & 1 \cr 0 & 1 & 0
\end{pmatrix}
\quad
\lX_7 =   \begin{pmatrix}
 0 & 0 & 0 \cr 0 & 0 & -i \cr 0 & i & 0
\end{pmatrix}
.
\EQ
Under $\SU(2).\U(1)$, these matrices have the same quantum numbers as
the scalar Higgs doublet of the standard model,
and they match the well known generators of $\SU(3)$.
However, they do not close by commutations on the $\lX_a$,
essentially because $Tr(\lX_0) \neq 0$~\eqref{eq28}. Rather~\eqref{eq25}, they close by anticommutation
\BQ  \label{eq36}
   \{\lX_i,\;\lX_j\}_+ = d^a_{ij}\;\lX_a\qquad a = 0,1,2,3\quad i,j = 4,5,6,7.
\EQ
Observe also that they transform left leptons into right leptons and vice-versa,
and therefore~\eqref{eq24} anticommute with the chirality operator
\BE  \label{eq37}
   [\chi, \lX_a]_- = \{\chi,\;\lX_i\}_+ = 0\qquad a = 0,1,2,3;\quad i = 4,5,6,7.
\EE
In this sense, the $\lX_a$ matrices are even, the $\lX_i$ are odd. Together
they define the fundamental irreducible representation of the superalgebra $\SU(2/1)$,
which appears first in Kac's classification~\cite{Kac1} under the name $A(1/0)$.

With respect to the super-Killing metric~\eqref{eq27} of $\SU(2/1)$
\BQ  \label{eq38}
g_{AB} = \frac {1}{2} \; STr (\lX_A\lX_B)\qquad A,B = 0,1\ldots 7
\EQ
the even subspace has a Minkowski signature $(-,+,+,+)$ and the electric charge operator
\BQ  \label{eq39}
\gamma =  -\frac {1}{2}\;\{\lX_6,\lX_6\}_+ =  \frac {1}{2}\; (\lX_0 + \lX_3) =
\begin{pmatrix}
 0 & 0 & 0 \cr 0 & -1 & 0 \cr 0 & 0 & -1
\end{pmatrix}
\EQ
is on the light cone $STr (\gamma^2) = 0$. In the terminology of Kac, the lepton
representation is atypical and its two Casimir operators~\eqref{eq30} vanish:
\BE  \label{eq40}
K_2 = K_3 = 0
\EE

This description of the $\SU(2/1)$ leptons was first proposed
independently by Ne'eman and Fairlie in 1979~\cite{N1,F1,NSF}.
Its most remarkable feature is that it unifies in a single irreducible
representation of the superalgebra the left and the right helicity state
of the electron. This would be impossible in a Lie algebra
where we would need to consider, as in the Georgi-Glashow grand-unified $\SU(5)$ model
the left anti-(right electron) $\overline{(e_R)}_L$.
A priori, the $\SU(2/1)$ symmetry could be exact at relatively low energy
whereas the $\SU(5)$ grand-unification scale is necessarily extremely high
to avoid a fast decay of particles into lighter antiparticles
and avoid a contradiction with the observed stability of the proton.

\section{The SU(2/1) antilepton representation}\label{A3}

Except for $\lX_0$, the lepton matrices~\eqref{eq31},~\eqref{eq33},~\eqref{eq35} look very familiar:
they coincide with those of the fundamental 3 dimensional
representation of $\SU(3)$. In particular, they are Hermitian.
But this is a coincidence. If we turn to the antilepton representation,
the electric charge of the positron is positive.
As we must maintain the definition of the photon~\eqref{eq39}, the odd matrices must be anti-Hermitian. In the basis
antielectron antineutrino: $((\overline{e_R})_L\;/ (\overline{e_L})_R, (\overline{\nu_L})_R))$
the even matrices matrix read
\BQ  \label{eq41}
\lX_0 =   \begin{pmatrix}
 2 & 0 & 0 \cr 0 & 1 & 0 \cr 0 & 0 & 1
\end{pmatrix},\quad
\lX_1 =   \begin{pmatrix}
 0 & 0 & 0 \cr 0 & 0 & 1 \cr 0 & 1 & 0
\end{pmatrix},\quad
\lX_2 =   \begin{pmatrix}
 0 & 0 & 0 \cr 0 & 0 & -i \cr 0 & i & 0
\end{pmatrix},\quad
\lX_3 =   \begin{pmatrix}
 0 & 0 & 0 \cr 0 & 1 & 0 \cr 0 & 0 & -1
\end{pmatrix},\quad
\EQ
and the 4 odd matrices read:
\BQ  \label{eq42}
\lX_4 =   \begin{pmatrix}
 0 & 0 & -1 \cr 0 & 0 & 0 \cr 1 & 0 & 0
\end{pmatrix},\quad
\lX_5 =   \begin{pmatrix}
 0 & 0 & i \cr 0 & 0 & 0 \cr i & 0 & 0
\end{pmatrix},\quad
\lX_6 =   \begin{pmatrix}
 0 & 1 & 0 \cr -1 & 0 & 0 \cr 0 & 0 & 0
\end{pmatrix},\quad
\lX_7 =   \begin{pmatrix}
 0 & -i & 0 \cr -i & 0 & 0 \cr 0 & 0 & 0
\end{pmatrix}\,.
\EQ
The sign of Killing metric~\eqref{eq27} flips because we must flip the chirality $\chi$, but all the structure constants~\eqref{eq25} are unchanged
and the electric charge~\eqref{eq39} of the positron reads
\BE  \label{eq43}
  \gamma = - \frac{1}{2}\{\lX_6,\;\lX_6\} = - \frac{1}{2}\{\lX_7,\;\lX_7\} = \frac{1}{2}(\lX_0 + \lX_3) =
\begin{pmatrix}
  1 & 0 & 0 \cr 0 & 1 & 0 \cr 0 & 0 & 0
\end{pmatrix}
\EE
The antilepton representation is also atypical, its two Casimir operators~\eqref{eq30} again vanish:
\BE  \label{eq44}
K_2 = K_3 = 0\,.
\EE

Probably because they are anti-Hermitian, the odd matrices of the
antilepton representation were never written explicitly in the $\SU(2/1)$ literature~\cite{NSF}, although a complete
theory of the anti-Fermions must be equivalent to a complete theory of the Fermions. Including both as in~\cite{CQ0}
would be over-counting. In reality, one cannot avoid facing these anti-Hermitian matrices since, as we
shall see in the next section, the odd quark matrices are partly Hermitian and partly ant-Hermitian.

\section{The SU(2/1) quark representation}\label{A4}

Let us now consider the up and down quarks $(u_R\;/\;(u_L,d_L)\;/\;d_R)$. They consist of an $\SU(2)$ left doublet and
two right singlets with known weak hyper-charges $4/3,1/3/1/3,-2/3$ and we can immediately write the even matrices.
\begin{align}
\label{eq45}
\chi &= \begin{pmatrix}
 -1 & 0 & 0 & 0 \cr 0 & 1 & 0 & 0 \cr 0 & 0 & 1 & 0 \cr 0 & 0 & 0 & -1
\end{pmatrix}
,\!\!\!\! &  \lX_0 &=
\begin{pmatrix}
 4/3 & 0 & 0 & 0 \cr 0 & 1/3 & 0 & 0 \cr 0 & 0 & 1/3 & 0 \cr 0 & 0 & 0 & -2/3
\end{pmatrix}
. \\
\lX_1 &=   \begin{pmatrix}
 0 & 0 & 0 & 0 \cr 0 & 0 & 1 & 0 \cr 0 & 1 & 0 & 0 \cr 0 & 0 & 0 & 0
\end{pmatrix} &
\lX_2 &=   \begin{pmatrix}
 0 & 0 & 0 & 0 \cr 0 & 0 & -i & 0 \cr 0 & i & 0 & 0 \cr 0 & 0 & 0 & 0
\end{pmatrix} \qquad\qquad
\lX_3 =   \begin{pmatrix}
 0 & 0 & 0 & 0 \cr 0 & 1 & 0 & 0 \cr 0 & 0 & -1 & 0 \cr 0 & 0 & 0 & 0
\end{pmatrix}
  \label{eq46}
\end{align}

It seemed a priori difficult to fit these 4 dimensional matrices in the $\SU(2/1)$ framework and the quarks
were listed in the original article of Ne'eman~\cite{N1} as a counterexample
to the $\SU(2/1)$ paradigm and left out in~\cite{F1}. But soon after, in what could be described as the first success of the model,
it was realized~\cite{DJ,NTM1} that such a representation had been found earlier by Scheunert, Nahm and Rittenberg~\cite{SNR}.
The existence of the 4 dimensional quark representation is natural considering the isomorphism of the superalgebras
$\SU(2/1)$ and $\OSp(2/2)$ which generalizes the well know Lie algebra isomorphisms
of $\SU(2)$, $\Sp(2)$ and $\SO(3)$.
The construction is simple. Since the electric charges of the up and down quarks ($u$ and $d$)
are respectively (2/3) and (-1/3), we can infer
from the definition~\eqref{eq39} of the photon matrix $\gamma = - \lX_6^2$ the form of $\lX_6$ and the other odd
matrices follow by commutation with $\SU(2)$. One obtains
\BQA   \label{eq47}
\lX_4 &= \frac {1}{\st} \begin{pmatrix}
  0 & 0 & -\sd & 0 \cr 0 & 0 & 0 & 1 \cr \sd & 0 & 0 & 0  \cr 0 & 1 & 0 & 0
\end{pmatrix} &
\lX_5 &= \frac {1}{\st}  \begin{pmatrix}
  0 & 0 & i\sd & 0 \cr 0 & 0 & 0 & -i \cr i\sd & 0 & 0 & 0  \cr 0 & i & 0 & 0
\end{pmatrix}
\\
\lX_6 &=  \frac {1}{\st}  \begin{pmatrix}
 0 & \sd & 0 & 0 \cr -\sd & 0 & 0 & 0 \cr 0 & 0 & 0 & 1 \cr 0 & 0 & 1 & 0
\end{pmatrix} &
\lX_7 &=  \frac {1}{\st}  \begin{pmatrix}
 0 & -i\sd & 0 & 0 \cr -i\sd & 0 & 0 & 0 \cr 0 & 0 & 0 & -i \cr 0 & 0 & i & 0
\end{pmatrix}
\EQA
A direct calculation shows that the quark matrices have the same commutators and anticommutators
as the lepton matrices.
In particular, we recognize the electric charge of the quarks in the diagonal photon matrix~\eqref{eq39}:
\BE\label{eq48}
  \gamma = - \frac{1}{2}\{\lX_6,\;\lX_6\} = - \frac{1}{2}\{\lX_7,\;\lX_7\} = \frac{1}{2}(\lX_0 + \lX_3) =
\begin{pmatrix}
  2/3 & 0 & 0 & 0\cr 0 & 2/3 & 0 &0 \cr 0 & 0 & -1/3 & 0 \cr 0 & 0 & 0 & -1/3
\end{pmatrix}
\EE
In Kac terminology, the quark representation is typical. Its Casimir operators~\eqref{eq30} are diagonal with eigenvalues
\BE   \label{eq49}
K_2 = \frac{8}{9}\;I \,,\quad K_3 = - \frac{64}{27}\;I.
\EE
As in the case of the antileptons, the quark matrices are never given explicitly in the literature.

\section{The OSp(2/2) neutral representation}\label{A5}

It must be stressed that $\SU(2/1)$ does not predict
that the hyper-charge of the left quark doublet is $1/3$.
This charge is a free parameter. One can construct
an irreducible representation of $\SU(2/1)$ of
arbitrary hyper-charge $1/n$ ($n$ can be a complex number) using
the same matrices $\chi$, $\lX_1$, $\lX_2$, $\lX_3$ as before,
selecting the desired values in $\lX_0$ and writing the corresponding $\lX_6$
\BQ   \label{eq50}
\lX_0 =\!
\begin{pmatrix}
 (n\!+\!1)/n & 0 & 0 & 0 \cr 0 & 1/n & 0 & 0 \cr 0 & 0 & 1/n & 0 \cr 0 & 0 & 0 & -(n\!-\!1)/n
\end{pmatrix}
,\quad
\lX_6 = \frac {1}{\sqrt {2n}}\!
\begin{pmatrix}
 0 & \sqrt{n\!+\!1} & 0 & 0 \cr -\sqrt{n\!+\!1} & 0 & 0 & 0 \cr 0 & 0 & 0 & \sqrt{n\!-\!1} \cr 0 & 0 & \sqrt{n\!-\!1} & 0
\end{pmatrix} .
\EQ
The other odd generators are constructed by commutation with the even $\lX_a$
and have the same shape as for the quarks.
When $n = \pm 1$, one of the 4 states (i.e.\ the right neutrino) decouples.
When $n = -1$, we recover the lepton representation,
when $n = 1$, the antileptons,
when $n = 3$, the quarks.
when $n = -3$, the antiquarks.
The conclusion is that in $\SU(2/1)$ the electric charge is
not quantized, but if the electric charge of the electron $e^-$ is equal to the
electric charge of the $\SU(2)$ lowering operator $\lX_1 - i\,\lX_2$,
i.e.\ to the charge of the observed $W^-$ Yang-Mills vector Boson,
then the right neutrino decouples,
it has no electric charge and no weak hyper-charge.
 Otherwise, in a quark like
representation where the weak hyper-charge of the $\SU(2)$ doublet is neither $1$ nor $-1$,
the 2 right singlets must exist, and their electric charge must differ by 1 unit.

In the large $n$ limit, we obtain the natural $\OSp(2/2)$ symmetric representation for which the doublet is $\SU(2)$ neutral.
\BQ   \label{eq51}
\lX_0 =
\begin{pmatrix}
 1 & 0 & 0 & 0 \cr 0 & 0& 0 & 0 \cr 0 & 0 & 0 & 0 \cr 0 & 0 & 0 & -1
\end{pmatrix}
\,,\quad
\lX_6 = \frac {1}{\sqrt 2}
\begin{pmatrix}
 0 & 1 & 0 & 0 \cr -1 & 0 & 0 & 0 \cr 0 & 0 & 0 & 1\cr 0 & 0 & 1 & 0
\end{pmatrix}
.
\EQ

In Kac terminology, the neutral representation is typical, its cubic super-Casimir operator~\eqref{eq30} vanishes:
\BE   \label{eq52}
K_2 = I \,,\quad K_3 = 0\,.
\EE

If we change variables and label the quadruplet representation by the hypercharge $y$ of the $\SU(2)$ doublet (y = 1/n of equation~\eqref{eq50})
we can label a family of representation by a vector $\{y_i\}$ giving the collection of its hypercharges. The standard
model family (electron + 3 quarks) is labeled by the vector $\{-1,1/3,1/3,1/3\}$. By definition,
the Adler triangle  anomaly $\U(1)\SU(2)\SU(2)$ cancels out if $\Sigma y_i = 0$.
In each representation $Y =  \text{diagonal} (y+1,y,y,y-1)$, hence $STr(Y^3)= -6y$ and the $\U(1)^3$ triangle anomaly also cancels out if   $\Sigma y_i = 0$.
By inspection, the scalar anomalies~\eqref{eq8} and~\eqref{eq12} are also proportional $\Sigma y_i$. We do not have a simple analytic proof
but verified numerically that $\Delta(\rho)$ of equation~\eqref{eq16} is linear in $y$. We conclude that any family such that  $\Sigma y_i = 0$
is anomaly free. We have already presented three examples. The standard model family (electron + 3 quarks), the anti-family
(positron + 3 antiquarks, the neutral family of Minahan, Ramond and Warner (a single $\OSp(2/2)$ neutral quark~\cite{MRW}).
But a model with one quark with $y=2/3$ and two quarks with $y=-1/3$ is also
anomaly free. The electric charges of the $d_R$ states would be $(y-1)/2$, i.e.\ (-1/6, -2/3, -2/3).

\section{The H/K Hermitian couplings}\label{A6}

For completeness we give here explicitly the $\mu_i^{\pm}$ matrices which
define the couplings of the $H$ and $K$ fields to the Fermions~\cite{HS,HPS}. In the
lepton representation, the $\mu_i$ matrices are Hermitian, so
$\mu^-_i = \lX_i$ and $\mu^+_i = 0 $. In the positively charged antilepton
representation, it is the opposite, $\mu^-_i =  0$ and
\BQ  \label{eq53}
\mu^+_4 =   \begin{pmatrix}
 0 & 0 & i \cr 0 & 0 & 0 \cr -i & 0 & 0
\end{pmatrix},\quad
\mu^+_5 =   \begin{pmatrix}
 0 & 0 & 1 \cr 0 & 0 & 0 \cr 1 & 0 & 0
\end{pmatrix},\quad
\mu^+_6 =   \begin{pmatrix}
 0 & -i & 0 \cr i & 0 & 0 \cr 0 & 0 & 0
\end{pmatrix},\quad
\mu^+_7 =   \begin{pmatrix}
 0 & -1 & 0 \cr -1 & 0 & 0 \cr 0 & 0 & 0
\end{pmatrix}\,.
\EQ
Finally, in the quark representation, the $\mu^-$ matrices coupled to the $H$ field
read
\BQA   \label{eq54}
\mu^-_4 &= \frac {1}{\st} \begin{pmatrix}
  0 & 0 & 0 & 0 \cr 0 & 0 & 0 & 1 \cr 0 & 0 & 0 & 0  \cr 0 & 1 & 0 & 0
\end{pmatrix} &
\mu^-_5 &= \frac {1}{\st}  \begin{pmatrix}
  0 & 0 & 0 & 0 \cr 0 & 0 & 0 & -i \cr 0 & 0 & 0 & 0  \cr 0 & i & 0 & 0
\end{pmatrix}
\\[4pt]
\mu^-_6 &=  \frac {1}{\st}  \begin{pmatrix}
 0 & 0 & 0 & 0 \cr 0 & 0 & 0 & 0 \cr 0 & 0 & 0 & 1 \cr 0 & 0 & 1 & 0
\end{pmatrix} &
\mu^-_7 &=  \frac {1}{\st}  \begin{pmatrix}
 0 & 0 & 0 & 0 \cr 0 & 0 & 0 & 0 \cr 0 & 0 & 0 & -i \cr 0 & 0 & i & 0
\end{pmatrix}
\EQA
and
the $\mu^+$ matrices coupled to the $K$ field read
\BQA   \label{eq55}
\mu^+_4 &= \frac {1}{\st} \begin{pmatrix}
  0 & 0 & i\sd & 0 \cr 0 & 0 & 0 & 0 \cr -i\sd & 0 & 0 & 0  \cr 0 & 0 & 0 & 0
\end{pmatrix} &
\mu^+_5 &= \frac {1}{\st}  \begin{pmatrix}
  0 & 0 & \sd & 0 \cr 0 & 0 & 0 & 0 \cr \sd & 0 & 0 & 0  \cr 0 & 0 & 0 & 0
\end{pmatrix}
\\[4pt]
\mu^+_6 &=  \frac {1}{\st}  \begin{pmatrix}
 0 & -i\sd & 0 & 0 \cr i\sd & 0 & 0 & 0 \cr 0 & 0 & 0 & 0 \cr 0 & 0 & 0 & 0
\end{pmatrix} &
\mu^+_7 &=  \frac {1}{\st}  \begin{pmatrix}
 0 & -\sd & 0 & 0 \cr -\sd & 0 & 0 & 0 \cr 0 & 0 & 0 & 0 \cr 0 & 0 & 0 & 0
\end{pmatrix}
\EQA
The $\mu^{\pm}_i$ matrices of the quarks are proportional to the non-zero $\mu^{\pm}_i$
of the leptons and antileptons, complemented by an extra line-column of zeroes, and
the commutators with the even matrices $[\lX_a,X_i] = f^j_{ai}X_j$ have the same
$f^j_{ai}$ structure constants when $X_i=\lX_i,\mu^+_i,\mu^-_i$ in any representation.

\section{The massive neutrino  SU(2/1) indecomposable representation}\label{A7}

The representations presented so far are irreducible. This means that
all the states belonging to such a representation are equivalent in the sense
that under the action of the superalgebra each of them generates all of them,
or in other words each orbit covers the whole representation.
In a Lie algebra, all finite dimensional representations are fully reducible.
This means that they can be written as block diagonal matrices,
where each block corresponds to an irreducible representation.
But in a superalgebra, some finite dimensional representations are indecomposable.
This means that the matrices are triangular, or in other words
some orbits do not cover the whole representation.
Rather than sketching the complete theory~\cite{Marcu,GQS,Yucai}, we construct
a few examples of $\SU(2/1)$ indecomposable representations relevant to the
classifications of the elementary particles~\cite{CQ0,HS,HPS}.

The simplest example is applicable to neutrinos.
Consider~\eqref{eq50} with $n = -1$, and let us call the four states $\nu_R /(\nu_L,e_L)/e_R$.
The right neutrino, $\nu_R$ is $\SU(2)$ and $\U(1)$ neutral. This is experimentally
correct, but we know that the neutrino has a very small but non-zero mass.
Contrary to the case of a Lie algebra, it is possible in $\SU(2/1)$ to add a small scalar couplings
of order $\alpha$ as follows.
\begin{align}
\nn
\chi &= \begin{pmatrix}
 -1 & 0 & 0 & 0 \cr 0 & 1 & 0 & 0 \cr 0 & 0 & 1 & 0 \cr 0 & 0 & 0 & -1
\end{pmatrix}
, & \lX_0 &=
\begin{pmatrix}
 0 & 0 & 0 & 0 \cr 0 & -1 & 0 & 0 \cr 0 & 0 & -1 & 0 \cr 0 & 0 & 0 & -2
\end{pmatrix}
\\[4pt]     \label{eq56}
\lX_1 &=   \begin{pmatrix}
 0 & 0 & 0 & 0 \cr 0 & 0 & 1 & 0 \cr 0 & 1 & 0 & 0 \cr 0 & 0 & 0 & 0
\end{pmatrix} &
\lX_2 &=   \begin{pmatrix}
 0 & 0 & 0 & 0 \cr 0 & 0 & -i & 0 \cr 0 & i & 0 & 0 \cr 0 & 0 & 0 & 0
\end{pmatrix} &
\lX_3 &=   \begin{pmatrix}
 0 & 0 & 0 & 0 \cr 0 & 1 & 0 & 0 \cr 0 & 0 & -1 & 0 \cr 0 & 0 & 0 & 0
\end{pmatrix}
\\[4pt]
\lX_4 &=  \begin{pmatrix}
  0 & 0 & 0 & 0 \cr 0 & 0 & 0 & 1 \cr \alpha & 0 & 0 & 0  \cr 0 & 1 & 0 & 0
\end{pmatrix}&
\lX_5 &=  \begin{pmatrix}
  0 & 0 & 0 & 0 \cr 0 & 0 & 0 & -i \cr i\alpha & 0 & 0 & 0  \cr 0 & i & 0 & 0
\end{pmatrix}&
\lX_6 &=  \begin{pmatrix}
 0 & 0 & 0 & 0 \cr -\alpha & 0 & 0 & 0 \cr 0 & 0 & 0 & 1 \cr 0 & 0 & 1 & 0
\end{pmatrix} &
\lX_7 &=  \begin{pmatrix}
 0 & 0 & 0 & 0 \cr -i\alpha & 0 & 0 & 0 \cr 0 & 0 & 0 & -i \cr 0 & 0 & i & 0
\end{pmatrix} \nn
\end{align}
A direct calculation shows that these modified matrices have the same commutators and anticommutators
as the lepton matrices.
The even matrices are equivalent to~\eqref{eq31},~\eqref{eq33} with an additional first line and first column of zeroes,
meaning that $\nu_R$ remains $\SU(2)\U(1)$ neutral. The last line and last column of the odd matrices
reproduce~\eqref{eq35}. The new $\alpha$ terms occur in the first column, but are omitted from the first
line. Thus, each matrix is block triangular. There are two highest vectors annihilated by all the raising operators
$\nu_R$ and $\nu_L$. Since $(\lX_6 - i \lX_7) \nu_R = - 2 \alpha \nu_L$, the orbit of $\nu_R$ is the whole representations
whereas the orbits of the three other states does not cover $\nu_R$. Notice that their is a single
free parameter, because the four terms in $\alpha$
are linked bu the action of $\SU(2)$. To verify that we still have a representation of the superalgebra,
we just need to compute one anticommutator, say $\{\lX_4,\lX_6\}$ and check that the lower left corner
element vanishes.
As shown in~\cite{HPS}, if $H^6$ acquires a vacuum expectation value $h$, the neutrino acquires a mass
of order $\alpha h$.
A contrario, if we try to apply this mechanism to a Lie algebra and consider the same matrices,
we would need to compute the commutator $[\lX_4,\lX_6]$, and we would generate a non-zero
value in the lower-left corner, verifying on this simple example that we cannot construct
a four-dimensional indecomposable representation of $\SU(3)$.

\section{The three generations  SU(2/1) indecomposable representation}\label{A8}

In our second example, we show that in $\SU(2/1)$, it is is possible to mix
several copies of the same irreducible representation, explaining the existence
of the three generations of leptons and quarks labeled by the electron, the muon and the tau.
Relative to a Lie algebra, the peculiarity is that we can construct a representation
were the maximal commuting Cartan subalgebra cannot be diagonalized.
Consider the 8x8 block triangular matrices
\BQ    \label{eq57}
\LX_a =  \begin{pmatrix}
  \lX_a & 0 \cr \theta\, \lXB_a & \lX_a
  \end{pmatrix}\,,
\quad
\LX_i =  \begin{pmatrix}
  \lX_i & 0 \cr \theta\, \lXB_i & \lX_i
  \end{pmatrix}
\EQ
where $\theta$ is a arbitrary mixing angle which can be thought of as a parametrization of the Cabbibo angle~\cite{HS}.
The $\lX$ are the quark matrices given in equations~\eqref{eq47},~\eqref{eq48}, $\lXB_0 = 4\sqrt{2}/3\;Id$, where $Id$  is the $4\times 4$ identity matrix,
$\lXB_a = 0\,,\quad a=1,2,3$, and the off diagonal odd matrices read
\BQA    \label{eq58}
\lXB_4 &= \frac{1}{\sqrt 3} \begin{pmatrix}
  0 & 0 & -1 & 0 \cr 0 & 0 & 0 & -\sqrt 2 \cr 1 & 0 & 0 & 0  \cr 0 & -\sqrt 2 & 0 & 0
\end{pmatrix}
\,, &
\lXB_5 &=  \frac{1}{\sqrt 3} \begin{pmatrix}
  0 & 0 & i & 0 \cr 0 & 0 & 0 & i \sqrt 2 \cr i & 0 & 0 & 0  \cr 0 & -i\sqrt 2 & 0 & 0
\end{pmatrix}
\\[4pt]
\lXB_6 &= \frac{1}{\sqrt 3} \begin{pmatrix}
 0 & 1& 0 & 0 \cr -1 & 0 & 0 & 0  \cr 0 & 0 & 0 & -\sqrt2 \cr 0 & 0 & -\sqrt 2 & 0
\end{pmatrix}
\,, &
\lXB_7 &=  \frac{1}{\sqrt 3} \begin{pmatrix}
 0 & -i & 0 & 0 \cr -i & 0 & 0 & 0  \cr 0 & 0 & 0 & i\sqrt2 \cr 0 & 0 & -i\sqrt 2 & 0
\end{pmatrix}
\EQA
By inspection, one can verify that these eight matrices have the same commutators as the
quark matrices and thus form an eight dimensional indecomposable representation of $\SU(2/1)$.
This representation was first proposed, up to notations, in~\cite{CQ0}, and is implicit
in~\cite{Marcu,Yucai}, but is not included in~\cite{GQS}
who  only analyze the case where the Cartan subalgebra is diagonal. We are grateful to
Coquereaux, Quella, Schomerus and Sorba
for clarifying this point.

N.B.: Sorry, but in the version published in JHEP, the signs of the $\sqrt2$ terms in the $\lXB$ matrices have
been switched by mistake (8 occurrences), the correct signs are provided here, in this new version of the Arxiv preprint.

The construction can be extended to
three generation using $12\times 12$ block triangular matrices.
\BQ  \label{eq59}
\LX =   \begin{pmatrix}
  \lX & 0 & 0\cr\lXB & \lX & 0 \cr \nu & \lXB & \lX
\end{pmatrix}
\,,\quad
\nu_6 =
\frac{1}{2 \sqrt 3} \begin{pmatrix}
  0 &  \sqrt 2 & 0 & 0 \cr -\sqrt 2 & 0 & 0 & 0 \cr 0 & 0 & 0 & -5\cr 0 & 0 & -5 & 0
\end{pmatrix}
\EQ
Along the diagonal, we have three identical copies of the quark representation.
Just below the diagonal, we have 2 copies of the previous `Cabbibo' construction.
The $\nu_a$ matrices again vanish for $a=1,2,3$, and $\nu_0 = 2\, Id$ is again proportional to the identity.
The matrix $\nu_6$ is constrained. The matrices
$\nu_i,i=4,5,7$ are deduced from $\nu_6$ by commutation with the $\SU(2)$ generators.
One can then introduce a parametrization $\alpha, \beta, \gamma$ of the Cabbibo-Kobayashi-Maskawa mixing angles~\cite{HS},
\BQ   \label{eq60}
\LX =   \begin{pmatrix}
  \lX & 0 & 0\cr \alpha\, \lXB & \lX & 0 \cr \gamma \nu & \beta\, \lXB & \lX
\end{pmatrix}
\EQ
and solve two linear equations to adjust the scale of $\nu_0$ and the entries in $\nu_6$.

The construction cannot be extended to four generations, because there would be to many
constraints in the lower left corner.
In physics terminology, $\SU(2/1)$ can describe the scalar mixing of 3 generations
of quarks or leptons using a single indecomposable twelve dimensional representation
but does not allow a fourth generation.

\enl

\section{The Adler-Bell-Jackiw vector anomaly} \label{A9}

Having recognized that the smallest representation of the $\SU(2/1)$
superalgebra correctly
describes the quantum numbers of the electrons and the quarks,
we have solved the classic static classification problem.
We now consider the quantum dynamic problem and wonder if the adjoint representation
of $\SU(2/1)$ can describe the vectors Bosons of the standard model.
As usual, we associate a real Yang-Mills vector field $A^a_{\mu}$ to each even generator $\lX_a$,
and postulate that its couplings to the Fermion fields are given by the chiral Weyl-Dirac Lagrangian
\BQ   \label{eq61}
\LAG^{(A)}_{\psi} = \overline {(\psi_R)}_L\;\SG^{\mu}(\partial_{\mu} + A^a_{\mu}\;\lambda_a) \;\psi_R +
\overline {(\psi_L)}_R\;\SB^{\mu}(\partial_{\mu} + A^a_{\mu}\;\lambda_a)\; \psi_L
\EQ
where the spin-one Pauli matrices $\SG$ map the right spinors on the left spinors
and the $\SB$ matrices map the left spinors on the right spinors. In Minkowski
space they can be represented as:
\BQA   \label{eq62}
  \SG_0 &= \begin{pmatrix}
 1 & 0 \cr 0 & 1
\end{pmatrix}
, &  \SG_1 &= \begin{pmatrix}
 0 & 1 \cr 1 & 0
\end{pmatrix}
, & \SG_2 &= \begin{pmatrix}
 0 & -i \cr i & 0
\end{pmatrix}
, &  \SG_3 &= \begin{pmatrix}
 1 & 0 \cr 0 & -1
\end{pmatrix}
\\[4pt]
  \SB_0 &= \begin{pmatrix}
 -1 & 0 \cr 0 & -1
\end{pmatrix}
, &  \SB_1 &= \begin{pmatrix}
 0 & 1 \cr 1 & 0
\end{pmatrix}
, &  \SB_2 &= \begin{pmatrix}
 0 & -i \cr i & 0
\end{pmatrix}
, &  \SB_3 &= \begin{pmatrix}
 1 & 0 \cr 0 & -1
\end{pmatrix}
\EQA
They are Hermitian and satisfy the chiral Clifford-Weyl relations
\BQA   \label{eq63}
\SG_{\mu}\SB_{\nu} + \SG_{\nu}\SB_{\mu} &= 2 g_{\mu\nu} \;I_L
\\
\SB_{\mu}\SG_{\nu} + \SB_{\nu}\SG_{\mu} &= 2 g_{\mu\nu} \;I_R
\EQA
where $g_{\mu\nu} = g^{\mu\nu}$ denotes the diagonal Minkowski metric $(-1,1,1,1)$. Importantly, if we compute the
trace of the product of four $\SG$ matrices we find a tensor with mixed symmetry
\BQA   \label{eq64}
  Tr (\SG_{\mu}\SB_{\nu}\SG_{\rho}\SB_{\sigma}) &= 2 (g_{\mu\nu}g_{\rho\sigma} - g_{\mu\rho}g_{\nu\sigma} + g_{\mu\sigma}g_{\nu\rho}
  +  i\;\epsilon_{\mu\nu\rho\SG})
\\
  Tr (\SB_{\mu}\SG_{\nu}\SB_{\rho}\SG_{\sigma}) &= 2 (g_{\mu\nu}g_{\rho\sigma} - g_{\mu\rho}g_{\nu\sigma} + g_{\mu\sigma}g_{\nu\rho}
  - i\;\epsilon_{\mu\nu\rho\SG})
\EQA
where the $g$ terms are symmetric, and $\epsilon$ is fully antisymmetric in $\mu\nu\rho\sigma$ with $\epsilon_{0123} = 1$.

\newpage

We know that the Yang-Mills theory is multiplicatively renormalizable but in the presence of chiral Fermion
there is an obstruction visible in the evaluation of the self interaction of the vectors.
The classical vertex is given by the term cubic in $A$ inside $\LAG = -1/4 F^2_{\mu\nu}$.
It is proportional to $f_{abc}\; \;\partial_{\mu}A^a_{\nu}\;A^b_{\mu}A^c_{\nu}$.
We must verify that the divergent part of the one-loop quantum correction has the same tensorial
structure as the classical term, so that, following Feynman's prescription, the divergence can
be absorbed in a redefinition of the coupling constant. In a non-chiral Yang-Mills theory, this is
always true. But as found by Adler~\cite{Adler} and Bell-Jackiw~\cite{BJ}, there is a subtle problem in the presence of chiral
Fermions. The calculation is at the same time complicated and well known, so we only
present a few crucial points. Consider the four distinct diagrams with 3 external vectors $A^a_{\mu}A^b_{\nu}A^c_{\rho}$
hitting a chiral Fermion loop.

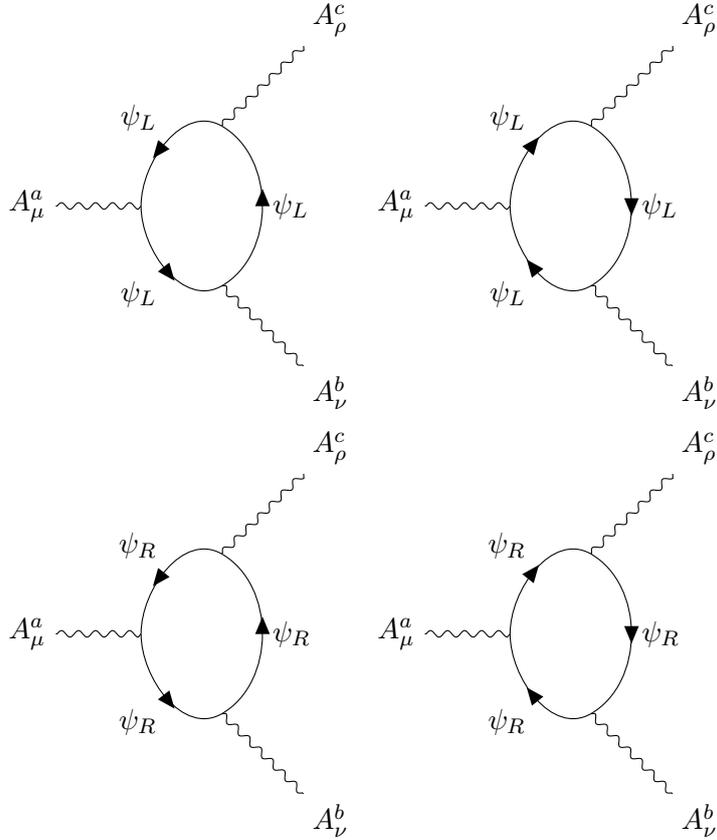
\begin{center}
\begin{tikzpicture}
\begin{feynman}
  \vertex (a1) {$A^a_{\mu}$};
\vertex [right= of a1] (a);
\vertex [below right=of a] (b);
\vertex [above right=of a] (c);
  \vertex [below right=of b](b1) {$A^b_{\nu}$};
  \vertex [above right=of c](c1) {$A^c_{\rho}$};
\diagram* {
  (a1) -- [photon] (a),
  (b1) -- [photon] (b),
  (c) -- [photon] (c1),
  (a) --  [anti fermion, in=150,out=90, edge label =$\psi_L$ ](c),
  (c) -- [anti fermion, in=30,out=-30, edge label =$\psi_L$ ] (b),
  (b) -- [anti fermion, in=270,out=210, edge label =$\psi_L$ ] (a),
};
\end{feynman}
\end{tikzpicture}
\begin{tikzpicture}
\begin{feynman}
  \vertex (a1) {$A^a_{\mu}$};
\vertex [right= of a1] (a);
\vertex [below right=of a] (b);
\vertex [above right=of a] (c);
  \vertex [below right=of b](b1) {$A^b_{\nu}$};
  \vertex [above right=of c](c1) {$A^c_{\rho}$};
\diagram* {
  (a1) -- [photon] (a),
  (b1) -- [photon] (b),
  (c) -- [photon] (c1),
  (a) --  [fermion, in=150,out=90, edge label =$\psi_L$ ](c),
  (c) -- [fermion, in=30,out=-30, edge label =$\psi_L$ ] (b),
  (b) -- [fermion, in=270,out=210, edge label =$\psi_L$ ] (a),
};
\end{feynman}
\end{tikzpicture}

\begin{tikzpicture}
\begin{feynman}
  \vertex (a1) {$A^a_{\mu}$};
\vertex [right= of a1] (a);
\vertex [below right=of a] (b);
\vertex [above right=of a] (c);
  \vertex [below right=of b](b1) {$A^b_{\nu}$};
  \vertex [above right=of c](c1) {$A^c_{\rho}$};
\diagram* {
  (a1) -- [photon] (a),
  (b1) -- [photon] (b),
  (c) -- [photon] (c1),
  (a) --  [anti fermion, in=150,out=90, edge label =$\psi_R$ ](c),
  (c) -- [anti fermion, in=30,out=-30, edge label =$\psi_R$ ] (b),
  (b) -- [anti fermion, in=270,out=210, edge label =$\psi_R$ ] (a),
};
\end{feynman}
\end{tikzpicture}
\begin{tikzpicture}
\begin{feynman}
  \vertex (a1) {$A^a_{\mu}$};
\vertex [right= of a1] (a);
\vertex [below right=of a] (b);
\vertex [above right=of a] (c);
  \vertex [below right=of b](b1) {$A^b_{\nu}$};
  \vertex [above right=of c](c1) {$A^c_{\rho}$};
\diagram* {
  (a1) -- [photon] (a),
  (b1) -- [photon] (b),
  (c) -- [photon] (c1),
  (a) --  [fermion, in=150,out=90, edge label =$\psi_R$ ](c),
  (c) -- [fermion, in=30,out=-30, edge label =$\psi_R$ ] (b),
  (b) -- [fermion, in=270,out=210, edge label =$\psi_R$ ] (a),
};
\end{feynman}
\end{tikzpicture}
\end{center}

First, each loop involves an integration over the momentum of the Fermions. The sign of the Fermion propagators
depends on the orientation of the loop. Since each diagram contains 3 propagators, the signs are flipped when
we reverse the orientation of the loops. The sum is antisymmetric under the simultaneous
exchange of $(b,\nu)$ with $(c,\rho)$.

Next, one must trace over the six Pauli matrices, 3 coming from the vertices, 3 from the propagators. For the left Fermion loops,
the propagators use the $\SB$ and the vertex use the $\SG$, and vice-versa for the right Fermion loops. Thus in~\eqref{eq64} the sign of the
$\epsilon_{\mu\nu\rho\SG}$ term depends on the chirality of each loop.

Finally, we trace over the $\lambda$ matrices. We get 2 terms. The first term from~\eqref{eq64} receives the same
sign from the left and right Fermions and is symmetric in $\nu\rho$ hence skew in $(bc)$ yielding $Tr (\lambda_a [\lambda_b,\lambda_c]_-)$.
In any representation of a Lie algebra, this trace~\eqref{eq28}
is proportional to the structure constants of the Lie algebra, as hoped for. However, the second term from~\eqref{eq64} with the $\epsilon$ Lorentz
structure is skew in $\nu\rho$ hence symmetric in $(bc)$ but as the overall sign depends on the chirality of the Fermion loop, this
term involves a supertrace. It has the wrong tensorial structure, should vanish, and its matrix dependent part reads
\BE   \label{eq65}
STr (\lambda_a \{\lambda_b,\lambda_c\}_+) = 0\,,
\EE
This term (2) is known as the triangle anomaly. It matches the even part~\eqref{eq29} of the cubic super-Casimir tensor of the $\SU(2/1)$ superalgebra.

As discovered by Bouchiat, Iliopoulos and Meyer~\cite{BIM} in 1972 for the group $\SU(2)\U(1)$, this term is non zero on the leptons
and the quarks but the sum of the 2 contributions vanishes when we have 3 quarks for every lepton. This implied the
existence of a pair of quark flavors for the electron (the up and down quarks), and a second pair (strange and charm)
associated to the muon at a time when the charm quark was not yet directly observed. It also implied the
existence of the top quark after the discovery of the $\tau$ lepton and bottom quark. For us, the occurrence of the
chiral supertrace indicates the need for a  chiral superalgebraic description of the electroweak interactions.

There is a second solution, discussed in the elegant note of Minahan, Ramond and Warner in 1990~\cite{MRW}, which also has a neat
$\SU(2/1)$ interpretation. In the large $n$ limit, the $\SU(2/1)$ neutral quarks of the previous section~\eqref{eq51}, with electric charge $\pm 1/2$,
have a vanishing $K_3$ Casimir operator~\eqref{eq30} and are anomaly free by themselves, hence they require no leptons.

Finally,  from the point of view of
$\SU(2/1)$, if we knew nothing of the strong interactions, we could also accept as a solution of the Adler-Bell-Jackiw constraints,
five $\SU(5)$ quarks with electric charge $3/5$ and $-2/5$ or more generally $n$ $\SU(n)$ quarks of electric-charge $(n+1)/2n$ and $-(n-1)/2n$.
In the $\SU(5)$ grand unified theory, which breaks down to $\SU(3)\SU(2)\U(1)$, the charge of the quarks are automatically $2/3$ and $-1/3$, but this is a
tautology, because if the strong interactions were described by an $\SU(5)$ group, we could as well have chosen an
$\SU(8)$ grand unified theory breaking down to $\SU(5)\SU(2)\U(1)$ and predict the charge of the quarks to be $3/5$ and $-2/5$.


\end{document}